%% file: main.tex
\documentclass[nonacm,sigconf]{acmart}

\input{style.tex}

\begin{document}
\title{MAD-DAG: Protecting Blockchain Consensus from MEV}

\author{Roi Bar-Zur}
\affiliation{%
  \institution{Technion}
  \city{}
  \state{}
  \country{Israel}
}

\author{Aviv Tamar}
\affiliation{%
  \institution{Technion}
  \city{}
  \state{}
  \country{Israel}
}

\author{Ittay Eyal}
\affiliation{%
  \institution{Technion}
  \city{}
  \state{}
  \country{Israel}
}

\begin{abstract}
    Blockchain security is threatened by \emph{selfish mining}, where a \emph{miner} (operator) deviates from the protocol to increase their revenue.
    Selfish mining is exacerbated by adverse conditions: \emph{rushing} (network propagation advantage for the selfish miner), \emph{varying block rewards} due to block contents, called \emph{miner extractable value} (MEV), and \emph{petty-compliant} miners who accept bribes from the selfish miner.
    
    The state-of-the-art selfish-mining-resistant blockchain protocol, \emph{Colordag}, does not treat these adverse conditions and was proven secure only when its latency is impractically high.

    We present \emph{MAD-DAG}, Mutual-Assured-Destruction Directed-Acyclic-Graph, the first practical protocol to counter selfish mining under adverse conditions.
    MAD-DAG achieves this thanks to its novel ledger function, which discards the contents of equal-length chains competing to be the longest.
    
    We analyze selfish mining in both Colordag and MAD-DAG by modeling a rational miner using a Markov Decision Process (MDP).
    We obtain a tractable model for both by developing conservative reward rules that favor the selfish miner to yield an upper bound on selfish mining revenue.
    To the best of our knowledge, this is the first tractable model of selfish mining in a practical DAG-based blockchain.
    This enables us to obtain a lower bound on the security threshold, the minimum fraction of computational power a miner needs in order to profit from selfish mining.

    MAD-DAG withstands adverse conditions under which Colordag and Bitcoin fail, while otherwise maintaining comparable security.
    For example, with petty-compliant miners and high levels of block reward variability, MAD-DAG's security threshold ranges from 11\% to 31\%, whereas both Colordag and Bitcoin achieve 0\% for all levels.
\end{abstract}

\maketitle

%%%%%%%%%%%%%%%%%%%%%%%%%%%%%%%%%%%%%%%%%%%%%%%%%%%%%%%%%%%%%%%%%%%%%%%%%%%%%%%%%%%%%%%%%%%%%%%%%%%%%%%%%%%%%%%%%%%%%%%%%%%%%%%%%%%%%%%%%%%%%%%%%%%%%%%%%%%%%%%%%%%%%%%%%%%%%%%%%%%%%%%%%%%%%%%%%%%%%%%%%%%%%%%%%%%%%%%%%%%%%%%%%%%%%%%%%%%%%%%%%%%%%%%%%%%%%%%%%%%%%%%%%%%%%%%%%%%%%%%%%%%%%%%%%%%%%%%%%%%%%%%%%%%%%%%%%%%%%%%%%%%%%%%%%%%%%%%%%%%%%%%%%%%%%%%%%%%%%%%%%%%%%%%%%%%%%%%%%%%%%%%%%%%%%%%%%%%%%%%%%%%%%%%%%%%%%%%%%%%%%%%%%%%%%%%%%%%%%%%%%%%%%%%%%%%%%%%%

\section{Introduction}

%%%%%%%%%%%%%%%%%%%%%%%%%%%%%%%%%%%%%%%%%%%%%%%%%%%%%%%%%%%%%%%%%%%%%%%%%%%%%%%%%%%%%%%%%%%%%%%%%%%%%%%%%%%%%%%%%%%%%%%%%%%%%%%%%%%%%%%%%%%%%%%%%%%%%%%%%%%%%%%%%%%%%%%%%%%%%%%%%%%%%%%%%%%%%%%%%%%%%%%%%%%%%%%%%%%%%%%%%%%%%%%%%%%%%%%%%%%%%%%%%%%%%%%%%%%%%%%%%%%%%%%%%%%%%%%%%%%%%%%%%%%%%%%%%%%%%%%%%%%%%%%%%%%%%%%%%%%%%%%%%%%%%%%%%%%%%%%%%%%%%%%%%%%%%%%%%%%%%%%%%%%%%%%%%%%%%%%%%%%%%%%%%%%%%%%%%%%%%%%%%%%%%%%%%%%%%%%%%%%%%%%%%%%%%%%%%%%%%%%%%%%%%%%%%%%%%%%%

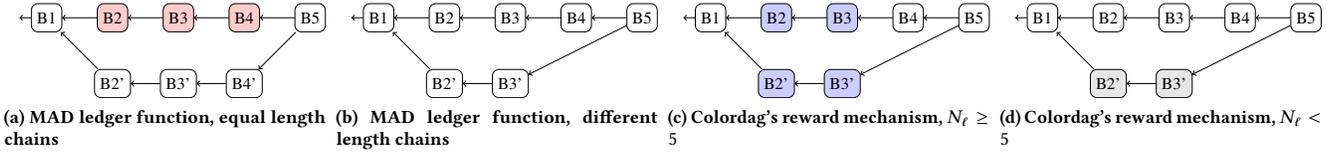
\begin{figure*}[!t]
    \centering
    \subfloat[MAD ledger function, equal length chains]{\centering
        \resizebox{0.24\textwidth}{!}{
        \begin{tikzpicture}[scale=1.4]
            \tikzstyle{block} = [draw, rectangle, rounded corners, minimum size=0.6cm]
            \tikzstyle{contested} = [draw, rectangle, rounded corners, minimum size=0.6cm, fill=red!20]
            \node (B0) at (-0.5, 0) {};
            \node[block] (B1) at (0, 0) {B1};
            \node[contested] (B2) at (1, 0) {B2};
            \node[contested] (B3) at (2, 0) {B3};
            \node[contested] (B4) at (3, 0) {B4};
            \node[block] (B5) at (4, 0) {B5};
            \node[block] (B2') at (1, -1) {B2'};
            \node[block] (B3') at (2, -1) {B3'};
            \node[block] (B4') at (3, -1) {B4'};

            \draw[->] (B1) -- (B0);
            \draw[->] (B2) -> (B1);
            \draw[->] (B3) -> (B2);
            \draw[->] (B4) -> (B3);
            \draw[->] (B5) -> (B4);
            \draw[->] (B2') -> (B1);
            \draw[->] (B3') -> (B2');
            \draw[->] (B4') -> (B3');
            \draw[->] (B5) -> (B4');
        \end{tikzpicture}}%
        \label{fig:mad-dag-ledger-function-1}}
    \hfil
    \subfloat[MAD ledger function, different length chains]{\centering
        \resizebox{0.24\textwidth}{!}{
        \begin{tikzpicture}[scale=1.4]
            \tikzstyle{block} = [draw, rectangle, rounded corners, minimum size=0.6cm]
            \tikzstyle{contested} = [draw, rectangle, rounded corners, minimum size=0.6cm, fill=red!20]
            \node (B0) at (-0.5, 0) {};
            \node[block] (B1) at (0, 0) {B1};
            \node[block] (B2) at (1, 0) {B2};
            \node[block] (B3) at (2, 0) {B3};
            \node[block] (B4) at (3, 0) {B4};
            \node[block] (B5) at (4, 0) {B5};
            \node[block] (B2') at (1, -1) {B2'};
            \node[block] (B3') at (2, -1) {B3'};

            \draw[->] (B1) -- (B0);
            \draw[->] (B2) -> (B1);
            \draw[->] (B3) -> (B2);
            \draw[->] (B4) -> (B3);
            \draw[->] (B5) -> (B4);
            \draw[->] (B2') -> (B1);
            \draw[->] (B3') -> (B2');
            \draw[->] (B5) -> (B3');
        \end{tikzpicture}}%
        \label{fig:mad-dag-ledger-function-2}}
    \hfil
    \subfloat[Colordag's reward mechanism, $\forkSensitivity \geq 5$]{\centering
        \resizebox{0.24\textwidth}{!}{
        \begin{tikzpicture}[scale=1.4]
            \tikzstyle{block} = [draw, rectangle, rounded corners, minimum size=0.6cm]
            \tikzstyle{contested} = [draw, rectangle, rounded corners, minimum size=0.6cm, fill=blue!20]
            \node (B0) at (-0.5, 0) {};
            \node[block] (B1) at (0, 0) {B1};
            \node[contested] (B2) at (1, 0) {B2};
            \node[contested] (B3) at (2, 0) {B3};
            \node[block] (B4) at (3, 0) {B4};
            \node[block] (B5) at (4, 0) {B5};
            \node[contested] (B2') at (1, -1) {B2'};
            \node[contested] (B3') at (2, -1) {B3'};

            \draw[->] (B1) -- (B0);
            \draw[->] (B2) -> (B1);
            \draw[->] (B3) -> (B2);
            \draw[->] (B4) -> (B3);
            \draw[->] (B5) -> (B4);
            \draw[->] (B2') -> (B1);
            \draw[->] (B3') -> (B2');
            \draw[->] (B5) -> (B3');
        \end{tikzpicture}}%
        \label{fig:colordag-reward-mechanism-1}}
    \hfil
    \subfloat[Colordag's reward mechanism, $\forkSensitivity < 5$]{\centering
        \resizebox{0.24\textwidth}{!}{
        \begin{tikzpicture}[scale=1.4]
            \tikzstyle{block} = [draw, rectangle, rounded corners, minimum size=0.6cm]
            \tikzstyle{unrewarded} = [draw, rectangle, rounded corners, minimum size=0.6cm, fill=gray!20]
            \node (B0) at (-0.5, 0) {};
            \node[block] (B1) at (0, 0) {B1};
            \node[block] (B2) at (1, 0) {B2};
            \node[block] (B3) at (2, 0) {B3};
            \node[block] (B4) at (3, 0) {B4};
            \node[block] (B5) at (4, 0) {B5};
            \node[unrewarded] (B2') at (1, -1) {B2'};
            \node[unrewarded] (B3') at (2, -1) {B3'};

            \draw[->] (B1) -- (B0);
            \draw[->] (B2) -> (B1);
            \draw[->] (B3) -> (B2);
            \draw[->] (B4) -> (B3);
            \draw[->] (B5) -> (B4);
            \draw[->] (B2') -> (B1);
            \draw[->] (B3') -> (B2');
            \draw[->] (B5) -> (B3');
        \end{tikzpicture}}%
        \label{fig:colordag-reward-mechanism-2}}
    \caption{Colordag's reward mechanism and MAD-DAG's ledger function. Red blocks are \emph{destructed}. Blue blocks are \emph{contested}. Gray blocks are \emph{unacceptable}.}
    \label{fig:colordag-and-mad-dag}
    \Description{Illustrates Colordag's reward mechanism and MAD-DAG's ledger function.}
\end{figure*}

%%%%%%%%%%%%%%%%%%%%%%%%%%%%%%%%%%%%%%%%%%%%%%%%%%%%%%%%%%%%%%%%%%%%%%%%%%%%%%%%%%%%%%%%%%%%%%%%%%%%%%%%%%%%%%%%%%%%%%%%%%%%%%%%%%%%%%%%%%%%%%%%%%%%%%%%

% Blockchain
Cryptocurrencies such as Bitcoin~\cite{nakamoto2008bitcoin} have become increasingly popular, with a market capitalization surpassing~\$3.5 trillion as of~2025~\cite{coinmarketcap2025}.
At their core lie \emph{blockchains}, decentralized protocols for maintaining an append-only ledger of blocks, each containing the hash of the previous block and a list of \emph{transactions}, transfers of assets among users and more elaborate logic.

% Proof of Work
In \emph{Proof of Work} (PoW) blockchains such as Bitcoin~\cite{nakamoto2008bitcoin}, Kaspa~\cite{sompolinsky2021phantom}, Litecoin~\cite{litecoin2024}, and Bitcoin Cash~\cite{bitcoincash2024}, \emph{miners} add new blocks by expending computational power to solve cryptographic puzzles~\cite{dwork1992pricing} and broadcasting the solutions.
In \emph{Nakamoto Consensus} (NC), the protocol underlying Bitcoin~\cite{nakamoto2008bitcoin}, miners extend the \emph{longest chain}\footnote{The chain that most work was expended on.} they observe.
For each block in the longest chain, miners receive two rewards: a \emph{subsidy} of newly created coins and the \emph{transaction fees} paid by users whose transactions are included.

% Selfish mining
Previous work~(\quickSectionRef{sec:related-work}) has identified that NC is vulnerable to \emph{selfish mining}~\cite{eyal2018majority,nayak2016stubborn,gervais2016security,sapirshtein2017optimal,hou2019squirrl,zur2020efficient,bar2022werlman}, where a miner controlling sufficient computational power can withhold blocks to build a private chain, then release it while it is longest, causing other miners' blocks to be discarded due to no longer being in the longest chain.
By doing so, the selfish miner increases their revenue.
The attack can be strengthened due to \emph{rushing}, where a well-connected miner propagates blocks faster to win ties, and due to \emph{petty-compliant miners}~\cite{carlsten2016instability,bar2023deep}, who can be bribed to \emph{mine on} (extend) the selfish miner's chain during ties.
These adverse conditions reduce the \emph{security threshold} (the minimum power needed for profitable selfish mining) from~25\%~\cite{sapirshtein2017optimal,zur2020efficient} to as low as~0\%~\cite{sapirshtein2017optimal,eyal2018majority,bar2023deep}.

% Varying block rewards
Even without rushing or petty compliance, security degrades under \emph{varying block rewards}.
These arise from \emph{whale transactions} with exceptionally high fees~\cite{liao2017incentivizing,bar2022werlman, sarenche2024bitcoin} or \emph{Maximal Extractable Value (MEV)}, where miners reorder or insert transactions to capture opportunities such as arbitrage~\cite{daian2020flash,qin2022quantifying}.
For Bitcoin in particular, as block subsidies halve every~4 years, miner revenue increasingly depends on varying fees~\cite{nakamoto2008bitcoin}.
In addition, projects like BitVM~\cite{aumayr2024bitvm,kolobov2025collidervm} that aim to enable smart contracts on Bitcoin will further increase reward variability.

% Colordag
FruitChains~\cite{pass2017fruitchains} and Colordag~\cite{abraham2023colordag} mitigate selfish mining under idealized assumptions: limited rushing, no petty-compliant miners, and constant block rewards.
Neither protocol was analyzed under adverse conditions that break these assumptions.
But even under idealized assumptions, FruitChains suffers from a selfish mining strategy that is always slightly profitable;
and Colordag's analysis showed its security threshold approaches~50\%, but only under a parameter choice that also yields impractically high~\emph{protocol latency}, that is, the delay before a block is unlikely to be discarded or denied subsidy (Colordag employs a mechanism that denies the subsidy of a subset of blocks to disincentivize selfish mining).
This leaves an open challenge to design a protocol that strictly disincentivizes selfish mining under adverse conditions and with practical latency.

% MAD-DAG
To this end, we introduce MAD-DAG~(\quickSectionRef{sec:mad-dag}), a DAG-based blockchain protocol with practical latency that, compared to both Colordag and NC, better resists selfish mining, especially under the aforementioned adverse conditions.
MAD-DAG achieves this thanks to its novel \emph{ledger function}.
A blockchain's ledger function determines the content of the \emph{ledger} (the transactions that are considered valid) given all the observed blocks and their contents.
The \emph{MAD (Mutual Assured Destruction)} ledger function takes a novel approach by completely discarding the content of blocks on the longest chain that have another chain with the same length (red blocks, Figure~\ref{fig:mad-dag-ledger-function-1}).
This ensures that the miner does not receive any  transactions fees or MEV.
This approach is in contrast to both NC and Colordag, which simply take the content of all blocks in the longest chain.
The MAD ledger function severely hinders a selfish miner's ability to create an alternative chain, since they would need it to be strictly longer, as ties yield contentless blocks.
This negates any advantage gained from rushing or from bribing petty-compliant miners.

To disincentivize selfish mining due to transaction fees, it could have sufficed to burn transactions fees and still consider the transactions themselves as valid.
But, this approach does not work for MEV, which is transparent to the consensus protocol, and a stronger measure is necessary, motivating our choice of the MAD ledger function.

Our MAD ledger function is also incentive compatible to petty-compliant miners:
If a selfish miner attempts to bribe a petty-compliant miner to mine on their chain that has equal length to the longest chain, the petty-compliant miner can point to both chains, causing their contents to be discarded and enabling the petty-compliant miner to use the content in their own block.

To evaluate MAD-DAG's security, we conduct the first analysis of selfish mining in DAG-based blockchains under adverse conditions~(\quickSectionRef{sec:analyzing-selfish-mining}).
We analyze both Colordag and MAD-DAG by modeling the mining process from the viewpoint of a single rational miner maximizing their revenue using a \emph{Markov Decision Process (MDP)}.
We solve each MDP using the Probabilistic Termination Optimization (PTO) method~\cite{zur2020efficient} with dynamic programming~\cite{bertsekas2012dynamic}.

To obtain a tractable model of the protocols, we upper-bound the selfish miner's revenue by slightly modifying the protocols' subsidy mechanisms.
Colordag's subsidy mechanism, which we use in MAD-DAG as well, rewards only blocks on the longest chain that are \emph{uncontested}, meaning there are no \emph{competing blocks} at identical heights (distance from root) on other \emph{acceptable} chains (non-blue blocks, Figure~\ref{fig:colordag-reward-mechanism-1}).
To determine which chains are considered \emph{acceptable}, Colordag measures their symmetric difference from the longest chain (the set of blocks unique to each chain).
Chains differing by more than a parameter~$\forkSensitivity$ are considered not acceptable and are discarded before determining uncontested blocks (gray blocks, Figure~\ref{fig:colordag-reward-mechanism-2}).
For analysis, we use a modified subsidy mechanism where the selfish miner's blocks are always acceptable, and others' blocks are acceptable only under a stricter condition.
This simplification leads to many states being equivalent, resulting in a significantly smaller state space.

Our simplified approach allows us to analyze both Colordag and MAD-DAG under practical latency by setting~$\forkSensitivity$ to low values.
This is the first time Colordag's security is analyzed under practical latency, whereas previous analysis~\cite{abraham2023colordag} showed analytically that there exists a choice of~$\forkSensitivity$ that yields a high security threshold, but this choice quickly becomes impractically high.

% Analysis of difficulty-adjustment mechanism
We first focus on the effect of the \emph{difficulty-adjustment mechanism} on the security of the protocols~(\quickSectionRef{sec:difficulty-adjustment-mechanism}).
The difficulty-adjustment mechanism in PoW blockchains periodically adjusts the difficulty of the PoW puzzle to maintain a stable block creation rate, even as the total mining power on the network fluctuates.
MAD-DAG's difficulty-adjustment mechanism is identical to NC's; the difficulty is adjusted once a fixed number of blocks have been added to the longest chain.
In contrast, Colordag employs a similar mechanism that only considers the blocks that were rewarded to ensure that the rate of subsidy is stable.
This choice, however, slightly reduces the impact of the subsidy penalty, as unrewarded blocks are also not counted by the difficulty-adjustment mechanism, resulting in a slight decrease in difficulty.

% We consider three different \emph{tie-breaking rules}, which specify how to choose among multiple chains of equal length:
% (1)~\emph{First-heard} (as in NC): miners choose the first chain they hear about;
% (2)~\emph{Random}: miners break ties at random;
% or (3)~\emph{Worst-case}: miners always choose the selfish miner's chain (the best possible scenario for a selfish miner employing rushing and bribing petty-compliant miners).

MAD-DAG's difficulty-adjustment mechanism increases the security threshold from~25\% in NC and~34\% in Colordag to~39\% in MAD-DAG, in the presence of a rushing adversary and in a practical setting where~$\forkSensitivity = 25$.
We further observe that when the subsidy is significantly lower than transaction fees, Colordag's security threshold drops significantly lower than NC's, while MAD-DAG's drops but to a level that is still higher than NC's.

% Analysis of ledger function
Next, we compare the impact of the ledger function on the security of the protocols~(\quickSectionRef{sec:ledger-function}).
While both Colordag and MAD-DAG's security thresholds decrease as the level of block reward variability is present, MAD-DAG's ledger function is more robust.
When petty-compliant miners are present, Colordag's security threshold drops to~0\% at a mild level of block reward variability while MAD-DAG's security threshold remains unaffected by their presence and remains above~11\% even with extreme block reward variability.

% Conclusion
In conclusion~(\quickSectionRef{sec:conclusion}), our contributions are as follows:
\begin{itemize}
    \item MAD-DAG, a protocol with a novel ledger function that discards the content of equal-length chains competing to be the longest chain;
    \item MDP of selfish mining in Colordag and MAD-DAG;
    \item conservative reward rules that upper-bound selfish mining revenue in Colordag and MAD-DAG, leading to the first tractable MDP of selfish mining in DAG-based blockchains; and
    \item analysis of Colordag and MAD-DAG with practical latency and adverse conditions, showing MAD-DAG is the most secure protocol to date against selfish mining.
\end{itemize}

As Bitcoin continues to evolve, it becomes increasingly likely that its block rewards will become more volatile, underscoring the need for a more secure protocol that can withstand adverse conditions.
MAD-DAG is directly applicable to address this looming threat.

%%%%%%%%%%%%%%%%%%%%%%%%%%%%%%%%%%%%%%%%%%%%%%%%%%%%%%%%%%%%%%%%%%%%%%%%%%%%%%%%%%%%%%%%%%%%%%%%%%%%%%%%%%%%%%%%%%%%%%%%%%%%%%%%%%%%%%%%%%%%%%%%%%%%%%%%%%%%%%%%%%%%%%%%%%%%%%%%%%%%%%%%%%%%%%%%%%%%%%%%%%%%%%%%%%%%%%%%%%%%%%%%%%%%%%%%%%%%%%%%%%%%%%%%%%%%%%%%%%%%%%%%%%%%%%%%%%%%%%%%%%%%%%%%%%%%%%%%%%%%%%%%%%%%%%%%%%%%%%%%%%%%%%%%%%%%%%%%%%%%%%%%%%%%%%%%%%%%%%%%%%%%%%%%%%%%%%%%%%%%%%%%%%%%%%%%%%%%%%%%%%%%%%%%%%%%%%%%%%%%%%%%%%%%%%%%%%%%%%%%%%%%%%%%%%%%%%%%

\section{Related Work}
\label{sec:related-work}

%%%%%%%%%%%%%%%%%%%%%%%%%%%%%%%%%%%%%%%%%%%%%%%%%%%%%%%%%%%%%%%%%%%%%%%%%%%%%%%%%%%%%%%%%%%%%%%%%%%%%%%%%%%%%%%%%%%%%%%%%%%%%%%%%%%%%%%%%%%%%%%%%%%%%%%%%%%%%%%%%%%%%%%%%%%%%%%%%%%%%%%%%%%%%%%%%%%%%%%%%%%%%%%%%%%%%%%%%%%%%%%%%%%%%%%%%%%%%%%%%%%%%%%%%%%%%%%%%%%%%%%%%%%%%%%%%%%%%%%%%%%%%%%%%%%%%%%%%%%%%%%%%%%%%%%%%%%%%%%%%%%%%%%%%%%%%%%%%%%%%%%%%%%%%%%%%%%%%%%%%%%%%%%%%%%%%%%%%%%%%%%%%%%%%%%%%%%%%%%%%%%%%%%%%%%%%%%%%%%%%%%%%%%%%%%%%%%%%%%%%%%%%%%%%%%%%%%%

We begin by describing why we focus on PoW blockchains~(\quickSectionRef{sec:related-work:pow-blockchains}), and then present previous work on selfish mining and on mitigations for it under various conditions~(\quickSectionRef{sec:related-work:selfish-mining}).
We then describe how analyzing selfish mining is done~(\quickSectionRef{sec:related-work:analyzing-selfish-mining}) and, last, survey the use of DAG structures in other protocols~(\quickSectionRef{sec:related-work:dags}).

%%%%%%%%%%%%%%%%%%%%%%%%%%%%%%%%%%%%%%%%%%%%%%%%%%%%%%%%%%%%%%%%%%%%%%%%%%%%%%%%%%%%%%%%%%%%%%%%%%%%%%%%%%%%%%%%%%%%%%%%%%%%%%%%%%%%%%%%%%%%%%%%%%%%%%%%

\subsection{Proof-of-Work Blockchains}
\label{sec:related-work:pow-blockchains}

%%%%%%%%%%%%%%%%%%%%%%%%%%%%%%%%%%%%%%%%%%%%%%%%%%%%%%%%%%%%%%%%%%%%%%%%%%%%%%%%%%%%%%%%%%%%%%%%%%%%%%%%%%%%%%%%%%%%%%%%%%%%%%%%%%%%%%%%%%%%%%%%%%%%%%%%

While PoW has been criticized for energy consumption~\cite{de2019renewable}, leading to alternatives like Proof of Stake~\cite{king2012ppcoin,bentov2016cryptocurrencies} and Proof of Space~\cite{dziembowski2015proofs,cohen2019chia}, Bitcoin's continued market dominance~\cite{coinmarketcap2025} demonstrates PoW's enduring relevance.
There is ongoing research to make PoW more efficient~\cite{tsabary2022tuning,lasla2022green,mirkin2023sprints}.
We focus on addressing its vulnerability to selfish mining.

%%%%%%%%%%%%%%%%%%%%%%%%%%%%%%%%%%%%%%%%%%%%%%%%%%%%%%%%%%%%%%%%%%%%%%%%%%%%%%%%%%%%%%%%%%%%%%%%%%%%%%%%%%%%%%%%%%%%%%%%%%%%%%%%%%%%%%%%%%%%%%%%%%%%%%%%

\subsection{Selfish Mining and Mitigations}
\label{sec:related-work:selfish-mining}

%%%%%%%%%%%%%%%%%%%%%%%%%%%%%%%%%%%%%%%%%%%%%%%%%%%%%%%%%%%%%%%%%%%%%%%%%%%%%%%%%%%%%%%%%%%%%%%%%%%%%%%%%%%%%%%%%%%%%%%%%%%%%%%%%%%%%%%%%%%%%%%%%%%%%%%%

\subsubsection{Without Adverse Conditions}

Previous work analyzed selfish mining strategies in Nakamoto consensus in simplified settings (such as constant block rewards)~\cite{eyal2018majority,nayak2016stubborn,sapirshtein2017optimal,gervais2016security,grunspan2018profitability,finkbeiner2025sok}.
Keller~\cite{keller2025automated} also analyzes selfish mining in DAG-based PoW consensus protocols in simplified settings.
In contrast, we consider realistic  adverse conditions, namely rushing, varying block rewards and petty-compliant miners.

Several works propose mitigations for selfish mining.
FruitChains~\cite{pass2017fruitchains} provides upper bounds on selfish mining revenue but suffers from zero-risk slightly profitable deviations.
Other proposals like Bobtail~\cite{bissias2020bobtail} and StrongChain~\cite{szalachowski2019strongchain} that modify PoW mechanisms, or the DAG-based Tailstorm~\cite{keller2023tailstorm}, lack comprehensive analysis of general selfish mining strategies.
Sarenche et al.~\cite{sarenche2024selfish} suggested changing the difficulty-adjustment mechanism to consider blocks that are not in the longest chain, but this does not address petty-compliant miners who will then be incentivized to ignore these blocks to negate the effect.
Colordag~\cite{abraham2023colordag} was theoretically proven to resist selfish mining when its latency is impractically high and there are no adverse conditions.
In contrast to all these works, MAD-DAG mitigates selfish mining under adverse conditions and with practical latency.

%%%%%%%%%%%%%%%%%%%%%%%%%%%%%%%%%%%%%%%%%%%%%%%%%%%%%%%%%%%%%%%%%%%%%%%%%%%%%%%%%%%%%%%%%%%%%%%%%%%%%%%%%%%%%%%%%%%%%%%%%%%%%%%%%%%%%%%%%%%%%%%%%%%%%%%%

\subsubsection{Rushing}

Well-connected miners can perform rushing attacks by quickly propagating their blocks in the network, e.g., through relay networks~\cite{corallo2017bitcoin,gencer2018decentralization}.
With perfect connectivity, the security threshold of NC drops to 0\%.

Random tie-breaking was proposed~\cite{eyal2018majority} to mitigate rushing's impact in NC.
While this reduces the advantage of a well-connected selfish miner, it slightly lowers the threshold against other adversaries~\cite{sapirshtein2017optimal}.
MAD-DAG's ledger function provides inherent resistance to rushing without compromising security against weakly connected miners.

%%%%%%%%%%%%%%%%%%%%%%%%%%%%%%%%%%%%%%%%%%%%%%%%%%%%%%%%%%%%%%%%%%%%%%%%%%%%%%%%%%%%%%%%%%%%%%%%%%%%%%%%%%%%%%%%%%%%%%%%%%%%%%%%%%%%%%%%%%%%%%%%%%%%%%%%

\subsubsection{Reward Variability}

Block rewards vary significantly due to transaction fees and MEV~\cite{daian2020flash,qin2022quantifying}.
Carlsten et al.~\cite{carlsten2016instability} and Tsabary and Eyal~\cite{tsabary2018gap} showed that when there are only transaction fees, deviating from the protocol becomes profitable for any miner regardless of their mining power.
WeRLman~\cite{bar2022werlman, sarenche2024bitcoin} analyzed a middle ground with whale transactions alongside subsidy, showing that reward variability increases selfish mining profitability and decreases the security threshold.
These motivate our design of MAD-DAG, which disincentivizes selfish mining even when rewards vary.

Fee smoothing mechanisms~\cite{pass2017fruitchains,budinsky2023mitigating} redistribute fees across multiple blocks to reduce variability.
However, these cannot address MEV, which is not generally detectable, and they incentivize alternative payment channels~\cite{tsabary2022ledgerhedger} to bypass the mechanism.
In contrast, MAD-DAG's ledger function addresses fee variability and MEV without incentivizing alternative payment channels.

Unlike fee smoothing mechanisms, alternative transaction fee mechanisms~\cite{roughgarden2021transaction,chung2023foundations} address incentive compatibility in the context of a single block;
they do not consider their impact on selfish mining strategies.

%%%%%%%%%%%%%%%%%%%%%%%%%%%%%%%%%%%%%%%%%%%%%%%%%%%%%%%%%%%%%%%%%%%%%%%%%%%%%%%%%%%%%%%%%%%%%%%%%%%%%%%%%%%%%%%%%%%%%%%%%%%%%%%%%%%%%%%%%%%%%%%%%%%%%%%%

\subsubsection{Petty-Compliant Miners}

Carlsten et al.~\cite{carlsten2016instability} introduced petty-compliant miners who make undetectable deviations when profitable.
For instance, petty-compliant miners facing ties in the longest chain would choose the chain that leaves out the transactions with the highest fees.
Such behavior opens an avenue of implicit bribing, where a selfish miner deliberately leaves out transactions to incentivize petty-compliant miners to mine on their chain.
Carlsten et al.~\cite{carlsten2016instability} showed that when there is no subsidy, petty-compliant miners cause the security threshold to drop to~0\%.
Later work~\cite{bar2023deep,sarenche2025mining} extended the analysis to include the impact of subsidy, and showed that while the security threshold is positive, petty-compliant miners significantly lower it.

Yaish et al.~\cite{yaish2023uncle} discovered strategic timestamp manipulation in Ethereum, corroborating the possibility of petty-compliant miners.

To our knowledge, MAD-DAG is the first protocol to address vulnerability to selfish mining due to petty-compliant miners.

%%%%%%%%%%%%%%%%%%%%%%%%%%%%%%%%%%%%%%%%%%%%%%%%%%%%%%%%%%%%%%%%%%%%%%%%%%%%%%%%%%%%%%%%%%%%%%%%%%%%%%%%%%%%%%%%%%%%%%%%%%%%%%%%%%%%%%%%%%%%%%%%%%%%%%%%

\subsection{Analyzing Selfish Mining}
\label{sec:related-work:analyzing-selfish-mining}

%%%%%%%%%%%%%%%%%%%%%%%%%%%%%%%%%%%%%%%%%%%%%%%%%%%%%%%%%%%%%%%%%%%%%%%%%%%%%%%%%%%%%%%%%%%%%%%%%%%%%%%%%%%%%%%%%%%%%%%%%%%%%%%%%%%%%%%%%%%%%%%%%%%%%%%%

Sapirshtein et al.~\cite{sapirshtein2017optimal} developed a method to find the optimal selfish mining strategy by solving a sequence of related MDPs.
Bar-Zur et al.~\cite{zur2020efficient} developed an alternative method, Probabilistic Termination Optimization (PTO), which requires solving a single MDP.
Both used \emph{dynamic programming} to optimize the MDP.

Dynamic programming is intractable for large models, leading subsequent works to use \emph{deep reinforcement learning}~\cite{hou2019squirrl,bar2022werlman,bar2023deep}.
However, deep reinforcement learning does not guarantee convergence to the optimal policy, providing only a lower bound on revenue, leading to an upper bound on the security threshold.

In this paper, we bound the selfish miner's revenue from above to obtain a tractable model that can be solved using dynamic programming, guaranteeing optimality and yielding a lower bound on the security threshold.
This allows us to analyze both Colordag and MAD-DAG, and to prove MAD-DAG indeed resists selfish mining under adverse conditions.

%%%%%%%%%%%%%%%%%%%%%%%%%%%%%%%%%%%%%%%%%%%%%%%%%%%%%%%%%%%%%%%%%%%%%%%%%%%%%%%%%%%%%%%%%%%%%%%%%%%%%%%%%%%%%%%%%%%%%%%%%%%%%%%%%%%%%%%%%%%%%%%%%%%%%%%%

\subsection{DAG-Based Protocols}
\label{sec:related-work:dags}

%%%%%%%%%%%%%%%%%%%%%%%%%%%%%%%%%%%%%%%%%%%%%%%%%%%%%%%%%%%%%%%%%%%%%%%%%%%%%%%%%%%%%%%%%%%%%%%%%%%%%%%%%%%%%%%%%%%%%%%%%%%%%%%%%%%%%%%%%%%%%%%%%%%%%%%%

DAG structures have been extensively explored in consensus protocols~\cite{wang2023sok}.
Previous works consider a PoW blockchain in a setting where the rate of block creation is high relative to network delay, causing frequent \emph{forks} (two blocks pointing to the same previous block).
These suggested using a DAG structure to protect the blockchain from \emph{double-spending} (reversing confirmed transactions)~\cite{sompolinsky2015secure,bagaria2019prism,fitzi2018parallel,yu2020ohie}.
Other works consider a similar setting and use a DAG structure to allow blocks which are not in the longest chain to also contain transactions, allowing the ledger to grow faster~\cite{lewenberg2015inclusive,sompolinsky2016spectre,sompolinsky2021phantom,sompolinsky2022dag}.
While these works focus on improving performance or security, they do not analyze strategic mining behavior such as selfish mining.

DAGs also appear in fundamentally different consensus protocols not based on mining.
There are distributed protocols in which all participants are known and which are designed for Byzantine fault tolerance~\cite{keidar2021all,danezis2022narwhal,spiegelman2022bullshark,malkhi2023bbca}.
A notable example is IOTA~\cite{silvano2020iota}, which has users create blocks containing their own transactions rather than relying on miners.
These protocols use DAGs as a tool to reach consensus rather than to align participants' incentives, as we do.

%%%%%%%%%%%%%%%%%%%%%%%%%%%%%%%%%%%%%%%%%%%%%%%%%%%%%%%%%%%%%%%%%%%%%%%%%%%%%%%%%%%%%%%%%%%%%%%%%%%%%%%%%%%%%%%%%%%%%%%%%%%%%%%%%%%%%%%%%%%%%%%%%%%%%%%%%%%%%%%%%%%%%%%%%%%%%%%%%%%%%%%%%%%%%%%%%%%%%%%%%%%%%%%%%%%%%%%%%%%%%%%%%%%%%%%%%%%%%%%%%%%%%%%%%%%%%%%%%%%%%%%%%%%%%%%%%%%%%%%%%%%%%%%%%%%%%%%%%%%%%%%%%%%%%%%%%%%%%%%%%%%%%%%%%%%%%%%%%%%%%%%%%%%%%%%%%%%%%%%%%%%%%%%%%%%%%%%%%%%%%%%%%%%%%%%%%%%%%%%%%%%%%%%%%%%%%%%%%%%%%%%%%%%%%%%%%%%%%%%%%%%%%%%%%%%%%%%%

\section{MAD-DAG}
\label{sec:mad-dag}

%%%%%%%%%%%%%%%%%%%%%%%%%%%%%%%%%%%%%%%%%%%%%%%%%%%%%%%%%%%%%%%%%%%%%%%%%%%%%%%%%%%%%%%%%%%%%%%%%%%%%%%%%%%%%%%%%%%%%%%%%%%%%%%%%%%%%%%%%%%%%%%%%%%%%%%%%%%%%%%%%%%%%%%%%%%%%%%%%%%%%%%%%%%%%%%%%%%%%%%%%%%%%%%%%%%%%%%%%%%%%%%%%%%%%%%%%%%%%%%%%%%%%%%%%%%%%%%%%%%%%%%%%%%%%%%%%%%%%%%%%%%%%%%%%%%%%%%%%%%%%%%%%%%%%%%%%%%%%%%%%%%%%%%%%%%%%%%%%%%%%%%%%%%%%%%%%%%%%%%%%%%%%%%%%%%%%%%%%%%%%%%%%%%%%%%%%%%%%%%%%%%%%%%%%%%%%%%%%%%%%%%%%%%%%%%%%%%%%%%%%%%%%%%%%%%%%%%%

MAD-DAG is a PoW blockchain protocol with a DAG structure.
Each block contains a list of transactions, and one or more pointers to previous blocks, making it a DAG.
Each block also contains a solution to a PoW puzzle: a number (\emph{nonce}) such that the hash of its concatenation with the block's content is smaller than a target value.
This target is called the \emph{difficulty}.

\emph{Honest} miners follow the protocol, which prescribes the following.
A miner should create a new block with transactions that are not yet included in existing blocks and pointers to all leaves in the block DAG~(i.e., blocks that are not referenced by any other block).
The miner should then attempt to find a valid solution to the PoW puzzle.
If the miner finds a solution, they should immediately publish the new block.
Upon receiving a block, honest miners should add it to their local copy of the block DAG and attempt to mine on it and on all other remaining leaves.

Transaction and block propagation occur via an underlying peer-to-peer communication network.
We assume standard blockchain networking solutions and leave the network layer out of our design.

As in Colordag~\cite{abraham2023colordag}, a block that references both some other block and its ancestor at the same time is not valid.

MAD-DAG comprises three key mechanisms, which we describe next.

%%%%%%%%%%%%%%%%%%%%%%%%%%%%%%%%%%%%%%%%%%%%%%%%%%%%%%%%%%%%%%%%%%%%%%%%%%%%%%%%%%%%%%%%%%%%%%%%%%%%%%%%%%%%%%%%%%%%%%%%%%%%%%%%%%%%%%%%%%%%%%%%%%%%%%%%

\paragraph{Difficulty-Adjustment Mechanism}

A difficulty-adjustment mechanism determines how to update the PoW difficulty, and subsequently the block-creation rate.

MAD-DAG's difficulty-adjustment mechanism adjusts the difficulty of the Proof-of-Work puzzle every \emph{epoch}, that is, after a fixed number of blocks is added to the longest chain.
The difficulty is multiplied by the ratio between the desired epoch time and the time the last epoch took.
This maintains a stable block-creation rate even as the total mining power on the network fluctuates.

This mechanism is also used in NC~\cite{nakamoto2008bitcoin}.
In Bitcoin, on average, every 2 weeks, 2016 blocks are added to the longest chain, for example.

%%%%%%%%%%%%%%%%%%%%%%%%%%%%%%%%%%%%%%%%%%%%%%%%%%%%%%%%%%%%%%%%%%%%%%%%%%%%%%%%%%%%%%%%%%%%%%%%%%%%%%%%%%%%%%%%%%%%%%%%%%%%%%%%%%%%%%%%%%%%%%%%%%%%%%%%

\paragraph{Subsidy Mechanism}

A subsidy mechanism determines which blocks are rewarded with subsidy.

We follow Colordag's subsidy mechanism, which rewards only a subset of blocks.
The \emph{canonical chain} is the longest chain in the DAG.
In case of a tie, a protocol-defined tie-breaking rule determines the canonical chain.
We consider 2 variants of MAD-DAG with different tie-breaking rules: miners either have to pick the first chain they see or pick one randomly.

A block is considered \emph{acceptable} if it is in a chain with a symmetric difference from the canonical chain of at most~$\forkSensitivity$ blocks~(non-gray blocks, Figure~\ref{fig:colordag-reward-mechanism-1} and Figure~\ref{fig:colordag-reward-mechanism-2}).
We term this threshold the \emph{fork sensitivity}.
A block is considered \emph{uncontested} if it is the only block with its height (maximum distance from the root) that is acceptable (white blocks are uncontested and blue blocks and contested).
MAD-DAG's subsidy mechanism rewards only uncontested blocks on the canonical chain.

%%%%%%%%%%%%%%%%%%%%%%%%%%%%%%%%%%%%%%%%%%%%%%%%%%%%%%%%%%%%%%%%%%%%%%%%%%%%%%%%%%%%%%%%%%%%%%%%%%%%%%%%%%%%%%%%%%%%%%%%%%%%%%%%%%%%%%%%%%%%%%%%%%%%%%%%

\paragraph{Ledger Function}

A ledger function determines which blocks compose the ledger, namely, the ordered list of transactions that are considered valid.

MAD-DAG uses the novel MAD ledger function.
A block is considered \emph{destructed} if there are two chains with equal length to the canonical chain where one chain contains the block and the other does not.
The MAD ledger function considers the contents of only non-destructed blocks in the canonical chain.

When two competing chains following a fork have equal length, a new block pointing to both chains may be created, causing all previous blocks since the fork to become destructed (the new block is not destructed).
Since these blocks are destructed, the all transactions within these blocks are discarded, rolling back the state of ledger to its state before these blocks.
Note that transactions that were previously added to the destructed blocks may once again be added to the ledger starting from the new block.

%%%%%%%%%%%%%%%%%%%%%%%%%%%%%%%%%%%%%%%%%%%%%%%%%%%%%%%%%%%%%%%%%%%%%%%%%%%%%%%%%%%%%%%%%%%%%%%%%%%%%%%%%%%%%%%%%%%%%%%%%%%%%%%%%%%%%%%%%%%%%%%%%%%%%%%%%%%%%%%%%%%%%%%%%%%%%%%%%%%%%%%%%%%%%%%%%%%%%%%%%%%%%%%%%%%%%%%%%%%%%%%%%%%%%%%%%%%%%%%%%%%%%%%%%%%%%%%%%%%%%%%%%%%%%%%%%%%%%%%%%%%%%%%%%%%%%%%%%%%%%%%%%%%%%%%%%%%%%%%%%%%%%%%%%%%%%%%%%%%%%%%%%%%%%%%%%%%%%%%%%%%%%%%%%%%%%%%%%%%%%%%%%%%%%%%%%%%%%%%%%%%%%%%%%%%%%%%%%%%%%%%%%%%%%%%%%%%%%%%%%%%%%%%%%%%%%%%%

\section{Analyzing Selfish Mining}
\label{sec:analyzing-selfish-mining}

%%%%%%%%%%%%%%%%%%%%%%%%%%%%%%%%%%%%%%%%%%%%%%%%%%%%%%%%%%%%%%%%%%%%%%%%%%%%%%%%%%%%%%%%%%%%%%%%%%%%%%%%%%%%%%%%%%%%%%%%%%%%%%%%%%%%%%%%%%%%%%%%%%%%%%%%%%%%%%%%%%%%%%%%%%%%%%%%%%%%%%%%%%%%%%%%%%%%%%%%%%%%%%%%%%%%%%%%%%%%%%%%%%%%%%%%%%%%%%%%%%%%%%%%%%%%%%%%%%%%%%%%%%%%%%%%%%%%%%%%%%%%%%%%%%%%%%%%%%%%%%%%%%%%%%%%%%%%%%%%%%%%%%%%%%%%%%%%%%%%%%%%%%%%%%%%%%%%%%%%%%%%%%%%%%%%%%%%%%%%%%%%%%%%%%%%%%%%%%%%%%%%%%%%%%%%%%%%%%%%%%%%%%%%%%%%%%%%%%%%%%%%%%%%%%%%%%%%

We begin with preliminaries on Markov Decision Processes, selfish mining in an MDP and in NC~(\quickSectionRef{sec:analyzing-selfish-mining:preliminaries}).
We then present selfish mining models for both Colordag and MAD-DAG, starting with the full model~(\quickSectionRef{sec:analyzing-selfish-mining:mad-dag-model}), followed by a simplified model that bounds the selfish mining revenue from above~(\quickSectionRef{sec:analyzing-selfish-mining:upper-bound-model}), and afterward we describe how we solve the MDP~(\quickSectionRef{sec:analyzing-selfish-mining:solving-the-mdp}).

%%%%%%%%%%%%%%%%%%%%%%%%%%%%%%%%%%%%%%%%%%%%%%%%%%%%%%%%%%%%%%%%%%%%%%%%%%%%%%%%%%%%%%%%%%%%%%%%%%%%%%%%%%%%%%%%%%%%%%%%%%%%%%%%%%%%%%%%%%%%%%%%%%%%%%%%

\subsection{Preliminaries}
\label{sec:analyzing-selfish-mining:preliminaries}

%%%%%%%%%%%%%%%%%%%%%%%%%%%%%%%%%%%%%%%%%%%%%%%%%%%%%%%%%%%%%%%%%%%%%%%%%%%%%%%%%%%%%%%%%%%%%%%%%%%%%%%%%%%%%%%%%%%%%%%%%%%%%%%%%%%%%%%%%%%%%%%%%%%%%%%%

\subsubsection{Markov Decision Processes}

A \emph{Markov Decision Process (MDP)} is an iterative process where starting from state~$s_0$, in each step~$t$ an \emph{agent} takes an \emph{action}~$a_t$ in a \emph{state}~$s_t$, receives a \emph{reward}~$R_t$ and causes the state to stochastically transition to a new state~$s_{t+1}$.
Formally, an MDP is a tuple $\left( S, A, P, R \right)$ where:
\begin{enumerate}
    \item $S$ is a set of states~$s \in S$;
    \item $A$ is the set of actions~$a \in A$;
    \item $P: S \times A \times S \to [0, 1]$ is the transition probability function from a state~$s$ to a state~$s'$ given an action~$a$: $P(s, a, s') \in [0, 1]$ such that $\sum_{s' \in S} P(s, a, s') = 1$; and
    \item $R: S \times A \times S \to \mathbb{R}$ is the reward function for a transition from state~$s$ to a state~$s'$ given an action~$a$: $R(s, a, s') \in \mathbb{R}$ such that~$R_t = R(s_t, a_t, s_{t+1})$.
\end{enumerate}

An MDP is solved by finding a policy $\pi: S \to A$ that maximizes the expected \emph{utility}~$u(\pi)$.
The \emph{stochastic shortest path} is a special case of an MDP, where there is a terminal state~$s_T \in S$ such that no further actions can cause the state to transition to another state and no rewards are received in the terminal state~\cite{bertsekas2012dynamic}.
A common utility function in this case is the expected sum of the rewards received until termination time~$T$ (the first step in which the terminal state is reached):
\begin{equation}
    u(\pi) = \mathbb{E} \left[ \sum_{t=0}^{T} R_t \right] .
\end{equation}

Two popular approaches for solving MDPs are \emph{dynamic programming}, which includes \emph{value iteration} and \emph{policy iteration}~\cite{bertsekas2012dynamic}, and converges to the optimal policy; and \emph{deep reinforcement learning}~\cite{arulkumaran2017deep,hou2019squirrl,bar2022werlman}, which utilizes machine learning to allow larger state and actions spaces, but is not guaranteed to converge to the optimal policy.

%%%%%%%%%%%%%%%%%%%%%%%%%%%%%%%%%%%%%%%%%%%%%%%%%%%%%%%%%%%%%%%%%%%%%%%%%%%%%%%%%%%%%%%%%%%%%%%%%%%%%%%%%%%%%%%%%%%%%%%%%%%%%%%%%%%%%%%%%%%%%%%%%%%%%%%%

\subsubsection{Selfish Mining in an MDP}
\label{sec:analyzing-selfish-mining:preliminaries:selfish-mining-in-an-mdp}

Honest miners follow the protocol.
For example, in NC, they mine on the last block in the longest chain they observe and reveal their blocks immediately.
A selfish miner may strategically choose which block to mine on and when to reveal their blocks.

% Example scenario
Consider the following scenario that illustrates the attack where the block reward is constant.
A selfish miner mines a block and keeps it secret while attempting to mine another block extending it.
If the selfish miner succeeds, they possess a secret chain of two blocks.
Meanwhile, honest miners continue mining on the longest chain they know, and eventually one of them mines a new block.
The selfish miner then reveals their secret two-block chain, causing the honest miner's new block to be discarded while the selfish miner claims the rewards for both blocks in their chain.

% Benefit of selfish mining
This strategy does not increase the total rewards the selfish miner receives in this instance, as they still earn rewards for two blocks just as they would have by mining honestly.
The benefit arises from the wasted effort of honest miners, in combination with the difficulty-adjustment mechanism.
When this scenario repeats, fewer blocks are added to the longest chain within the epoch's target time period, causing the protocol to perceive a lower network block-creation rate.
Consequently, the difficulty decreases for subsequent epochs, enabling faster block creation that increases the selfish miner's revenue over time.
Hence, the utility function for a selfish miner is the rate at which rewards are received.

% Utility function for selfish mining
An MDP model for selfish mining needs to account for the difficulty-adjustment mechanism~\cite{eyal2018majority,sapirshtein2017optimal,zur2020efficient}.
To analyze selfish mining in an MDP, Bar-Zur et al.~\cite{zur2020efficient} suggest adding to the MDP another element: the \emph{difficulty contribution} function~$D: S \times A \times S \to \mathbb{R}^+$ that maps a transition from state~$s$ to state~$s'$ given an action~$a$ to how much the transition contributed towards the difficulty-adjustment mechanism such that~$D_t = D(s_t, a_t, s_{t+1})$.
The utility function of the agent would then be expected ratio between the total reward and the total difficulty contribution:
\begin{equation}
    u(\pi) = \mathbb{E} \left[ \frac{\sum_{t=0}^{\infty} R_t}{\sum_{t=0}^{\infty} D_t} \right] .
\end{equation}

While this is a non-standard utility function, Bar-Zur et al.~\cite{zur2020efficient} show that an MDP with such a utility function can be transformed to a stochastic shortest path MDP whose optimal policy is also with a bounded error to the optimal policy in the original MDP with the non-standard utility function.
The stochastic shortest path MDP can then be solved using standard techniques.

%%%%%%%%%%%%%%%%%%%%%%%%%%%%%%%%%%%%%%%%%%%%%%%%%%%%%%%%%%%%%%%%%%%%%%%%%%%%%%%%%%%%%%%%%%%%%%%%%%%%%%%%%%%%%%%%%%%%%%%%%%%%%%%%%%%%%%%%%%%%%%%%%%%%%%%%

\subsection{Colordag and MAD-DAG Model}
\label{sec:analyzing-selfish-mining:mad-dag-model}

%%%%%%%%%%%%%%%%%%%%%%%%%%%%%%%%%%%%%%%%%%%%%%%%%%%%%%%%%%%%%%%%%%%%%%%%%%%%%%%%%%%%%%%%%%%%%%%%%%%%%%%%%%%%%%%%%%%%%%%%%%%%%%%%%%%%%%%%%%%%%%%%%%%%%%%%

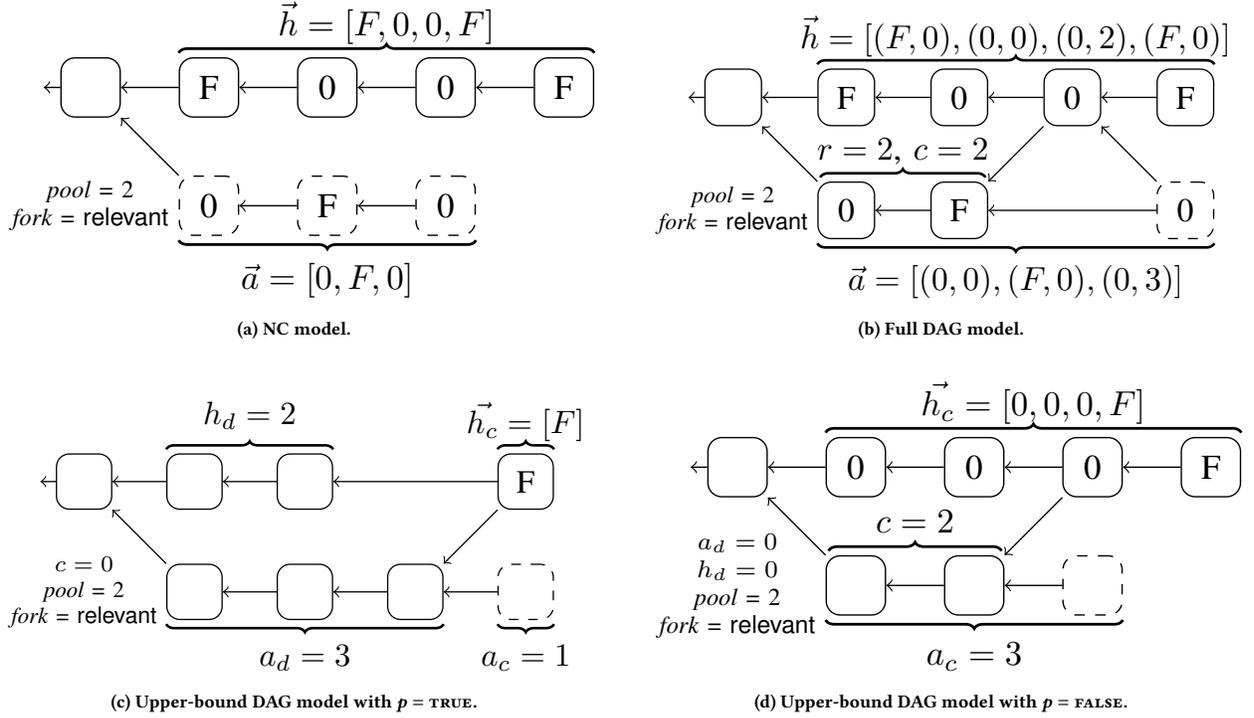
\begin{figure*}[t]
    \centering
    \subfloat[NC model.]{\resizebox{0.45\textwidth}{!}{
        \begin{tikzpicture}[scale=1.2]
            \tikzstyle{block} = [draw, rectangle, rounded corners, minimum size=0.6cm]
            \tikzstyle{secret} = [draw, rectangle, rounded corners, minimum size=0.6cm, dashed]

            \node (B0) at (-0.5, 0) {};
            \node[block] (B1) at (0, 0) {};
            \node[block] (B2) at (1, 0) {F};
            \node[block] (B3) at (2, 0) {0};
            \node[block] (B4) at (3, 0) {0};
            \node[block] (B5) at (4, 0) {F}; 
            \node[secret] (B2') at (1, -1) {0};
            \node[secret] (B3') at (2, -1) {F};
            \node[secret] (B4') at (3, -1) {0};

            \draw[->] (B1) -- (B0);
            \draw[->] (B2) -> (B1);
            \draw[->] (B3) -> (B2);
            \draw[->] (B4) -> (B3);
            \draw[->] (B5) -> (B4);
            \draw[->] (B2') -> (B1);
            \draw[->] (B3') -> (B2');
            \draw[->] (B4') -> (B3');

            \draw [thick,decoration={brace,mirror,raise=0.35cm},decorate] (B2'.west) -- (B4'.east) node [pos=0.5,anchor=north,yshift=-0.45cm] {$\aChain = [0, F, 0]$};

            \draw [thick,decoration={brace,raise=0.35cm},decorate] (B2.west) -- (B5.east) node [pos=0.5,anchor=north,yshift=1cm] {$\hChain = [F, 0, 0, F]$};

            \node [align=center,font=\scriptsize] (aux) at (0, -1) {\pool = 2 \\ \fork = \forkRelevant};
        \end{tikzpicture}}%
        \label{fig:state-space:nc}}
    \hfil
    \subfloat[Full DAG model.]{\resizebox{0.45\textwidth}{!}{
        \begin{tikzpicture}[scale=1.2]
            \tikzstyle{block} = [draw, rectangle, rounded corners, minimum size=0.6cm]
            \tikzstyle{secret} = [draw, rectangle, rounded corners, minimum size=0.6cm, dashed]

            \node (B0) at (-0.5, 0) {};
            \node[block] (B1) at (0, 0) {};
            \node[block] (B2) at (1, 0) {F};
            \node[block] (B3) at (2, 0) {0};
            \node[block] (B4) at (3, 0) {0};
            \node[block] (B5) at (4, 0) {F}; 
            \node[block] (B2') at (1, -1) {0};
            \node[block] (B3') at (2, -1) {F};
            \node[secret] (B4') at (4, -1) {0};

            \draw[->] (B1) -- (B0);
            \draw[->] (B2) -> (B1);
            \draw[->] (B3) -> (B2);
            \draw[->] (B4) -> (B3);
            \draw[->] (B5) -> (B4);
            \draw[->] (B2') -> (B1);
            \draw[->] (B3') -> (B2');
            \draw[->] (B4') -> (B3');

            \draw[->] (B4) -- (B3');
            \draw[->] (B4') -- (B4);

            \draw [thick,decoration={brace,mirror,raise=0.35cm},decorate] (B2'.west) -- (B4'.east) node [pos=0.5,anchor=north,yshift=-0.45cm] {$\aChain = [(0,0), (F,0), (0,3)]$};

            \draw [thick,decoration={brace,raise=0.35cm},decorate] (B2.west) -- (B5.east) node [pos=0.5,anchor=north,yshift=1cm] {$\hChain = [(F,0), (0,0), (0,2), (F,0)]$};

            \draw [thick,decoration={brace,raise=0.35cm},decorate] ($(B2'.west)+(0.025,0)$) -- ($(B3'.east)+(-0.025,0)$) node [pos=0.5,anchor=north,yshift=0.9cm,align=center] {$r = 2$, $c = 2$};

            \node [align=center,font=\scriptsize] (aux) at (0, -1) {\pool = 2 \\ \fork = \forkRelevant};
        \end{tikzpicture}}%
        \label{fig:state-space:full-dag}}
    
    \subfloat[Upper-bound DAG model with $\hap = \textsc{true}$.]{\resizebox{0.45\textwidth}{!}{
        \begin{tikzpicture}[scale=1.2]
            \tikzstyle{block} = [draw, rectangle, rounded corners, minimum size=0.6cm]
            \tikzstyle{secret} = [draw, rectangle, rounded corners, minimum size=0.6cm, dashed]

            \node (B0) at (-0.5, 0) {};
            \node[block] (B1) at (0, 0) {};
            \node[block] (B2) at (1, 0) {};
            \node[block] (B3) at (2, 0) {};
            \node[block] (B4) at (4, 0) {F}; 
            \node[block] (B2') at (1, -1) {};
            \node[block] (B3') at (2, -1) {};
            \node[block] (B4') at (3, -1) {};
            \node[secret] (B5') at (4, -1) {};

            \draw[->] (B1) -- (B0);
            \draw[->] (B2) -> (B1);
            \draw[->] (B3) -> (B2);
            \draw[->] (B4) -> (B3);
            \draw[->] (B2') -> (B1);
            \draw[->] (B3') -> (B2');
            \draw[->] (B4') -> (B3');
            \draw[->] (B5') -> (B4');

            \draw[->] (B4) -- (B4');

            \draw [thick,decoration={brace,mirror,raise=0.35cm},decorate] (B2'.west) -- (B4'.east) node [pos=0.5,anchor=north,yshift=-0.45cm] {$a_d = 3$};
            
            \draw [thick,decoration={brace,mirror,raise=0.35cm},decorate] (B5'.west) -- (B5'.east) node [pos=0.5,anchor=north,yshift=-0.45cm] {$a_c = 1$};
            
            \draw [thick,decoration={brace,raise=0.35cm},decorate] (B2.west) -- (B3.east) node [pos=0.5,anchor=north,yshift=1cm] {$h_d = 2$};
            
            \draw [thick,decoration={brace,raise=0.35cm},decorate] (B4.west) -- (B4.east) node [pos=0.5,anchor=north,yshift=1cm] {$\hChainAfter = [F]$};

            \node [align=center,font=\scriptsize] (aux) at (0, -1) {$c = 0$ \\ \pool = 2 \\ \fork = \forkRelevant};
        \end{tikzpicture}}%
        \label{fig:state-space:simplified-dag-1}}
    \hfil
    \subfloat[Upper-bound DAG model with $\hap = \textsc{false}$.]{\resizebox{0.45\textwidth}{!}{
        \begin{tikzpicture}[scale=1.2]
            \tikzstyle{block} = [draw, rectangle, rounded corners, minimum size=0.6cm]
            \tikzstyle{secret} = [draw, rectangle, rounded corners, minimum size=0.6cm, dashed]

            \node (B0) at (-0.5, 0) {};
            \node[block] (B1) at (0, 0) {};
            \node[block] (B2) at (1, 0) {0};
            \node[block] (B3) at (2, 0) {0};
            \node[block] (B4) at (3, 0) {0};
            \node[block] (B5) at (4, 0) {F}; 
            \node[block] (B2') at (1, -1) {};
            \node[block] (B3') at (2, -1) {};
            \node[secret] (B4') at (3, -1) {};

            \draw[->] (B1) -- (B0);
            \draw[->] (B2) -> (B1);
            \draw[->] (B3) -> (B2);
            \draw[->] (B4) -> (B3);
            \draw[->] (B5) -> (B4);
            \draw[->] (B2') -> (B1);
            \draw[->] (B3') -> (B2');
            \draw[->] (B4') -> (B3');

            \draw[->] (B4) -- (B3');

            \draw [thick,decoration={brace,mirror,raise=0.35cm},decorate] (B2'.west) -- (B4'.east) node [pos=0.5,anchor=north,yshift=-0.45cm] {$a_c = 3$};

            \draw [thick,decoration={brace,raise=0.35cm},decorate] (B2.west) -- (B5.east) node [pos=0.5,anchor=north,yshift=1cm] {$\hChainAfter = [0, 0, 0, F]$};

            \draw [thick,decoration={brace,raise=0.35cm},decorate] ($(B2'.west)+(0.025,0)$) -- ($(B3'.east)+(-0.025,0)$) node [pos=0.5,anchor=north,yshift=0.9cm,align=center] {$c = 2$};

            \node [align=center,font=\scriptsize] (aux) at (0, -1) {$a_d = 0$ \\ $h_d = 0$ \\ \pool = 2 \\ \fork = \forkRelevant};
        \end{tikzpicture}}%
        \label{fig:state-space:simplified-dag-2}}
    \caption{Example states in the NC and DAG models.}
    \label{fig:state-space}
    \Description{Example states in the NC and DAG models.}
\end{figure*}

We now present models of selfish mining in Colordag and MAD-DAG.
For brevity, we present the models in a unified manner, where different variants can be selected to represent the different protocols.

As in previous models~\cite{sapirshtein2017optimal,bar2022werlman}, we consider a single rational selfish miner and all other miners are honest.
The selfish miner has a fraction~$\alpha$ of the total mining power in the network, meaning that there is always a probability of~$\alpha$ that the selfish miner is the one who creates a new block.
The miner has a rushing factor~$\gamma$, meaning that when rushing, the selfish miner's block would reach a~$\gamma$-fraction of the other miners before the newly created block.
Except for the case of rushing, the network delay is zero:
When a miner publishes a block, all other miners instantly receive it.

Note that we only analyze the one-color version of Colordag since its multicolor mechanism is aimed to mitigate the effects of \emph{benign} forks (due to non-zero network delay) on the revenue of honest miners, while we assume the network delay is zero.

The subsidy is constant and worth~$1$ unit.
There occasionally appear whale transactions, each worth~$F$ units, and each block can contain at most one whale transaction.
Unlike previous models, we also assume that there is a constant source of transactions that can be included in every block and net a total of~$f$ units.
We term~$f$ the \emph{guaranteed fee}.

The selfish miner works only on a single secret chain at once, as in previous models~\cite{sapirshtein2017optimal,zur2020efficient,bar2022werlman}.
The selfish miner always references their own last block, since this never harms their revenue.
But they are allowed to also reference a block on the \emph{public chain}, the chain other miners are working on.
Note that they are allowed to reference up to one additional block from there as, as any additional one would result in pointing to both a block and to its ancestor, making the selfish miner's block invalid.

All other miners are honest:
They reference all known blocks and immediately publish blocks they create.

The model tracks the state of the selfish miner's and other miners' chains since the~\emph{last divergence}, that is, the last block that everyone agrees on whether it is in the canonical chain and whether it is acceptable, uncontested, or neither.

We now present the different variants of the model, followed by its utility function, its states, its actions, and its transitions and their yielded rewards and difficulty contributions~(the terms in the denominator of the utility function, see \quickSectionRef{sec:analyzing-selfish-mining:preliminaries:selfish-mining-in-an-mdp}).

%%%%%%%%%%%%%%%%%%%%%%%%%%%%%%%%%%%%%%%%%%%%%%%%%%%%%%%%%%%%%%%%%%%%%%%%%%%%%%%%%%%%%%%%%%%%%%%%%%%%%%%%%%%%%%%%%%%%%%%%%%%%%%%%%%%%%%%%%%%%%%%%%%%%%%%%

\subsubsection{Modes}

To allow us to analyze the effect of the different mechanisms of MAD-DAG and Colordag and of the adverse conditions, we present different variants of the different mechanisms we use.
Each combination of the following modes results with model for particular protocol.

\paragraph{Tie-Breaking}

The tie-breaking rule specifies how honest miners choose the canonical chain that determines subsidy rewards and transactions in the ledger.
Unlike classical blockchains, that only allow blocks to specify a single parent block, determining the canonical chain exactly, a DAG allows blocks to reference multiple parents, but the order in which they are referenced can determine the canonical chain (e.g., by preferring the block referenced first).

We consider three tie-breaking rule, which were analyzed before in the context of NC~\cite{eyal2018majority,sapirshtein2017optimal}.
We describe these rules in the context of our model, in the case that a selfish miner publishes an alternative chain of equal length to the chain other miners are working on, and they honest miners have to decide which chain to pick.
\begin{enumerate}
    \item First-heard: Honest miners choose the first chain they hear of.
    If the selfish miner had a block ready in advance, rushing is possible and the selfish miner's rushing factor determines how many honest miners choose the selfish miner's chain.
    \item Random: Honest miners choose a chain uniformly at random.
    The selfish miner's rushing factor does not affect how many honest miners choose the selfish miner's chain.
    \item Worst-case: All miners choose the selfish miner's chain.
    This mode provides an upper bound on the revenue of a selfish miner in the case that the miner has extreme rushing capability or that all other miners are petty-compliant.
\end{enumerate}

Note that a protocol can prescribe miners a specific tie-breaking rule like the first-heard rule or the random rule, but the worst-case rule is only used for analysis.

\paragraph{Difficulty-Adjustment Mechanism}

The difficulty-adjustment mechanism determines how the block-creation rate is updated every epoch.
In the context of the model, this determines the difficulty contributions.

We consider the following two difficulty-adjustment mechanisms.
\begin{enumerate}
    \item Uncontested: Only uncontested blocks contribute to the difficulty.
    \item Canonical: The difficulty is adjusted based on all blocks in the canonical chain.
\end{enumerate}

\paragraph{Ledger Function}
The ledger function determines the content of the ledger: which transactions are considered valid and net fees to the miner who included them.

We consider the two following ledger functions.
\begin{enumerate}
    \item Canonical: The ledger is the concatenation of all blocks in the canonical chain.
    \item MAD: The ledger is the concatenation of all non-destructed blocks in the canonical chain.
\end{enumerate}

Colordag uses the uncontested difficulty-adjustment mechanism.
While it doesn't specify its exact ledger function, the canonical ledger function is a natural choice.
MAD-DAG uses the canonical difficulty-adjustment mechanism and the MAD ledger function.

%%%%%%%%%%%%%%%%%%%%%%%%%%%%%%%%%%%%%%%%%%%%%%%%%%%%%%%%%%%%%%%%%%%%%%%%%%%%%%%%%%%%%%%%%%%%%%%%%%%%%%%%%%%%%%%%%%%%%%%%%%%%%%%%%%%%%%%%%%%%%%%%%%%%%%%%

\subsubsection{Utility Function}

Similarly to previous models~\cite{sapirshtein2017optimal,zur2020efficient}, the utility function is the expected rate of revenue of the selfish miner, the ratio of the total reward from subsidy, constant transactions, and whale transactions received to the total difficulty contribution based on the chosen difficulty-adjustment mechanism.

%%%%%%%%%%%%%%%%%%%%%%%%%%%%%%%%%%%%%%%%%%%%%%%%%%%%%%%%%%%%%%%%%%%%%%%%%%%%%%%%%%%%%%%%%%%%%%%%%%%%%%%%%%%%%%%%%%%%%%%%%%%%%%%%%%%%%%%%%%%%%%%%%%%%%%%%

\subsubsection{States}

A state in the model is defined by ${\left( \aChain, \hChain, \fork, r, c, \pool \right)}$, where~$\aChain$ is a vector of all the selfish miner's blocks since the last divergence (the last block up to which everyone agrees on the state of all mechanisms), each block contains a flag indicating whether the block contains a whale transaction and a reference (possibly null) to a block in the public chain;
$\hChain$ is the public chain, a vector of all honest miners' blocks since the last divergence with similar flags for transactions and references;
$\fork$ is a flag indicating whether rushing is possible (\forkRelevant) or not (\forkIrrelevant);
$r$ is the number of the selfish miner blocks that have been published;
$c$ is the index of the last selfish miner block in the canonical chain (only when there is ambiguity due to chains of equal length);
and $\pool$ is number whale transactions available since the last divergence.

Figure~\ref{fig:state-space:full-dag} shows an example state.

%%%%%%%%%%%%%%%%%%%%%%%%%%%%%%%%%%%%%%%%%%%%%%%%%%%%%%%%%%%%%%%%%%%%%%%%%%%%%%%%%%%%%%%%%%%%%%%%%%%%%%%%%%%%%%%%%%%%%%%%%%%%%%%%%%%%%%%%%%%%%%%%%%%%%%%%

\subsubsection{Actions and Transitions}

The actions in the model are as follows.

\paragraph{Reveal~$\ell$}
Reveal the~$\ell$-th miner's block~(${r \gets \ell}$).
Only possible if~$r < \ell$.

\paragraph{Mine~$\ell$}
The selfish miner attempts mining a block pointing to the last block in their chain and to the~$\ell$-th honest block ($\ell = 0$ corresponds to null).
The selfish miner succeeds with probability~$\alpha$.

Honest miners attempt mining a block pointing to the last block in the public chain and the last revealed selfish miner block (unless it was referenced in a previous honest block or there are no revealed selfish miner blocks).
If there are two possible candidates for the canonical chain, honest miners pick one of them based on the tie-breaking rule and the value of \fork.
If they pick the selfish miner's block, we update~$c$ accordingly.
The honest miners succeed with probability~${(1 - \alpha)}$.

If the creator of the new block is selfish (respectively, honest), the value of \fork is updated to \forkIrrelevant (\forkRelevant).

\paragraph{Merge}
The selfish miner discards all unrevealed blocks, accepts the current public chain as being the canonical one, and moves on to start a new secret chain extending it.
Both chains are reset to be empty~(${\aChain \gets \emptyChain}$ and ${\hChain \gets \emptyChain}$).
Reward and difficulty contribution are yielded based on the ledger function and the difficulty-adjustment mechanism, respectively, as follows.

First, we calculate the canonical chain based on the DAG structure of~$\aChain$ and~$\hChain$, breaking ambiguities due to tie-breaks with~$c$.
Then, we calculate which blocks are acceptable and which are uncontested, using the Colordag's algorithm~\cite{abraham2023colordag}.
After this is done, we can determine the subsidy the selfish miner receives and the difficulty contribution of the blocks in the canonical chain.
Then, if the ledger function is the MAD one, we calculate which blocks are destructed, by finding all chains with equal length to the canonical chain.
After this is done, we can determine the reward the selfish miner receives from constant fees for blocks in the canonical chain that the selfish miner created and from whale transactions if these blocks contain whale transactions.

For each possible transition, which results in a new block being added to the longest chain, we add a new transition where a whale transaction is added to the pool and everything else is the same.
The new transition has probability~$\delta$ times the probability of the original transition.
Then, we normalize all the transitions' probabilities to sum to~1.

This approach is different from previous ones~\cite{bar2022werlman,sarenche2024bitcoin}.
Sarenche et al.~\cite{sarenche2024bitcoin} argue that simply incrementing the pool, as Bar-Zur et al.~\cite{bar2022werlman} do, gives the selfish miner an unfair advantage by allowing them to predict in advance whether a new whale transaction become available until the next block is created.
Instead, they suggest to simultaneously add the whale transaction to the pool and to the new block that is being created.
This, however, restricts the miner's ability to react to the new whale transaction in time and change the block they are trying to mine.
We instead choose to interrupt the transition in the model when a new whale transaction is added to prevent the miner from knowing that a transaction will be available in advance but still give the selfish miner a chance to react to it to change their action.

\subsubsection{Bounding the State Space}

We allow the length of~$\aChain$ and~$\hChain$ to be up to a value of \maxFork.
To uphold this, the action Mine is disallowed if the length of either chain is equal to \maxFork.
While this caps the size of the state space, the model remains intractable for useful values of \maxFork.
For example, there are roughly 100,000,000 when \maxFork is 5.
To overcome this, we next present a model that uses an upper bound on the revenue of the selfish miner.

\subsubsection{Utility of Honest Mining}

If the selfish miner were to mine honestly, they would get a fraction of~$\alpha$ of all blocks.
Each block is worth~$1 + f + q F$, where~$q$ is the average number of whale transactions in a block.
Overall the utility of honest mining is then~$\alpha (1 + f + q F)$.

In previous models with fees~\cite{bar2022werlman,bar2023deep,sarenche2024bitcoin}, we simply have~$q=\delta$, as whenever a new whale transaction appears an honest miner would immediately include it.
But using the new approach to update~\pool, results with a positive possibility that more than one whale transaction appears before a new block is created.
Combined with the fact that transactions may overflow if there are more than~\maxPool simultaneously, it is no longer guaranteed that all whale transactions that appear will be included.

In our model, it holds that~${q = \frac{\delta - \delta^{\maxPool + 1}}{1 - \delta^{\maxPool + 1}}}$.
We defer the proof to Appendix~\ref{appendix:average-whales-included}.

We corroborate our results by simulating the honest policy and comparing the simulated revenue to the one we calculate with our closed-form solution.

%%%%%%%%%%%%%%%%%%%%%%%%%%%%%%%%%%%%%%%%%%%%%%%%%%%%%%%%%%%%%%%%%%%%%%%%%%%%%%%%%%%%%%%%%%%%%%%%%%%%%%%%%%%%%%%%%%%%%%%%%%%%%%%%%%%%%%%%%%%%%%%%%%%%%%%%

\subsection{Upper-Bound Model}
\label{sec:analyzing-selfish-mining:upper-bound-model}

%%%%%%%%%%%%%%%%%%%%%%%%%%%%%%%%%%%%%%%%%%%%%%%%%%%%%%%%%%%%%%%%%%%%%%%%%%%%%%%%%%%%%%%%%%%%%%%%%%%%%%%%%%%%%%%%%%%%%%%%%%%%%%%%%%%%%%%%%%%%%%%%%%%%%%%%

The upper-bound model is based on the full model presented in the previous section with several modifications that benefit the selfish miner.

First, we assume the selfish miner always includes as many whale transactions as they can, even if their blocks have been created before the transactions appeared.
Thanks to this, we do not need to keep track of the content of the selfish miner's chain and can only track its length.

Second, if the ledger function is the MAD one, when the selfish miner has at least as many blocks in their secret chain as the public chain, we allow the selfish miner to cause the blocks in the public chain to be destructed without the selfish miner having to reveal a chain of equal length to the public chain.
This only harms the revenue of honest miners and leaves more whale transactions to the selfish miner.

Third, we use a different notion for blocks to be acceptable:
A selfish miner's block is always acceptable;
an honest miner's block is acceptable if there is a path that contains it that has a symmetric difference from the canonical chain of at most~$\forkSensitivity$ blocks, and both the block's closest ancestor and closest descendant on the canonical chain were created by an honest miner.
The selfish miner having more acceptable blocks reduces the number of uncontested honest miners' blocks, and consequently benefit the selfish miner when the uncontested difficulty-adjustment mechanism is used.
The honest miners having less acceptable blocks increases the number of uncontested blocks the selfish miner has, and consequently benefit the selfish miner by increasing the reward they get.
While this also increases the difficulty contribution when the uncontested difficulty-adjustment mechanism is used, this overall revenue increases~($\frac{a+x}{b+x} > \frac{a}{b}$ when $b > a > 0$ and $x > 0$).
Thanks to the different notion we use, there is no longer any benefit for the selfish miner to reference blocks from the public chain, unless they are in the canonical chain (to allow the selfish miner to start a new secret chain) and in that case, it does not affect the acceptability of honest miners' blocks, and cannot lead to the selfish miner's blocks being penalized.
In addition, there is no longer any benefit for honest miners to reference blocks from the selfish miner's chain as any such chain would not count for the new notion of acceptability.

%%%%%%%%%%%%%%%%%%%%%%%%%%%%%%%%%%%%%%%%%%%%%%%%%%%%%%%%%%%%%%%%%%%%%%%%%%%%%%%%%%%%%%%%%%%%%%%%%%%%%%%%%%%%%%%%%%%%%%%%%%%%%%%%%%%%%%%%%%%%%%%%%%%%%%%%

\subsubsection{States}

As in the previous model, states track all blocks since the last divergence.
In the new model, we do so in a more structured manner~(Figure~\ref{fig:state-space:simplified-dag-1}).
The state tracks the blocks since the fork between the public chain~$\hChainAfter$ (including flags for whale transactions) and the selfish miner's secret chain~$a_c$.
With the canonical ledger function, the fork is also the last fork in the DAG, and these are all blocks that may end up in the canonical chain and may not.
With the MAD ledger function, this is not the case,
$\hChainAfter$ and~$a_c$ also track blocks for which it is known whether they are in the canonical chain or not, but it is not yet determined whether they are destructed or not.
Therefore, in this model, the element~$c$, the index of the last selfish miner block out of the~$a_c$ blocks in the secret chain, is only relevant for the MAD ledger function.

In addition, we track blocks that have been created prior to the fork between the public chain and the selfish miner's secret chain but after the last divergence (termed \emph{pre-fork} blocks).
These are blocks that everyone agrees whether they are part of the canonical chain, but it is not yet clear whether they are acceptable, uncontested, or neither.
No such block created by an honest miner can be in the canonical chain, because then it would be acceptable and everyone can determine whether it is uncontested (whether there is a selfish miner's block of the same height, as these are always acceptable), contradicting the way we selected these blocks.
In fact, the only case where such blocks may exist is when the selfish miner's blocks are in the canonical chain and the honest miners' blocks are not, and it is not yet clear whether the honest miners' blocks are acceptable.
In this case, they would become acceptable if and only if the current public chain would become the canonical chain.

Denote the number of pre-fork blocks that the selfish miner created by~$a_d$ and the number of pre-fork blocks that the honest miners created by~$h_d$.
If there are more than~$\forkSensitivity$ such blocks~(${a_d + h_d > \forkSensitivity}$), then the honest miners' blocks become unacceptable, as the only relevant path that contains these blocks has a symmetric difference to the canonical chain that is too large.
In that case, the selfish miner's blocks are uncontested and can be rewarded.
This also means that the divergence has been resolved (only partially, the canonical chain is not yet known).
These blocks are no longer relevant for the state.

However, at this stage, until an honest miner's block is added to the canonical chain, all honest miners' blocks that are not part of the canonical chain would also be unacceptable.
To track this, we introduce a new flag, the \emph{honest acceptability potential}~$\hap$, that indicates whether we are in this situation that honest miners' block out of the canonical chain have no potential to become acceptable~(Figure~\ref{fig:state-space:simplified-dag-2}).

Summarizing, a state is defined by ${\left( a_d, a_c, h_d, \hChainAfter, \fork, c, \pool, \hap \right)}$, where~$a_d$ is the number of pre-fork blocks that the selfish miner created;
$a_c$ is the number of blocks the selfish miner created in their secret chain;
$h_d$ is the number of pre-fork blocks that the honest miners created;
$\hChainAfter$ is a vector of all blocks in the public chain, the honest miners' blocks since the fork the selfish miner created (including flags for whale transactions);~$\fork$ is a ternary flag indicating whether rushing is possible (\forkRelevant) or not (\forkIrrelevant) or currently occurring (\forkActive);
$c$ is the index of the last selfish miner block (with respect to~$a_c$) that is in the canonical chain;
$\pool$ is the number of whale transactions available since the fork between the public chain and the selfish miner's secret chain;
and~$\hap$ is a flag indicating whether there is potential for all honest miners' blocks that are not part of the canonical chain to become acceptable.

We bound the state space similarly to how we do so in the full model.

%%%%%%%%%%%%%%%%%%%%%%%%%%%%%%%%%%%%%%%%%%%%%%%%%%%%%%%%%%%%%%%%%%%%%%%%%%%%%%%%%%%%%%%%%%%%%%%%%%%%%%%%%%%%%%%%%%%%%%%%%%%%%%%%%%%%%%%%%%%%%%%%%%%%%%%%

\subsubsection{Actions and Transitions}

The following actions are available.

\paragraph{Adopt~$\ell$}
The selfish miner discards their secret chain and adopts the first~$\ell$ blocks in~$\hChainAfter$.
This is only allowed if~$\ell \geq a_c$ (otherwise, the selfish miner should reveal their secret chain instead).

With the MAD ledger function, if~$c > 0$ and the number of blocks until the last fork in the DAG~(${a_d + c + h_d + c}$) is greater than~$\forkSensitivity$ then the honest miners' blocks that were discarded are not acceptable, making the selfish miner's blocks uncontested, yielding a reward of~$a_d + c$.
Otherwise, the honest blocks are acceptable, yielding a reward of~$a_d - h_d$.

The difficulty contribution is based on the chosen difficulty-adjustment mechanism.
If it is the uncontested difficulty-adjustment mechanism, a difficulty contribution of the reward plus~$\ell - a_c$ is yielded.
Otherwise, $a_d + \ell$ is yielded.

The public chain is shifted back by~$\ell$ blocks (${\hChainAfter \gets \hChainAfter \ll \ell}$) and the transactions that were in the discarded blocks are decremented from~$\pool$~(${\pool \gets \pool - \transactions{\hChainAfter[:\ell]}}$).
The selfish miner's secret chain and all other values are reset~(${a_c \gets 0}$, ${a_d \gets 0}$, ${h_d \gets 0}$, $\hap \gets \textsc{true}$).

\paragraph{Reveal~$\ell$}
The selfish miner reveals the first~$\ell$ blocks in their secret chain.
This is only allowed if~${\ell \geq \abs{\hChainAfter}}$.

If the selfish miner reveals a strictly longer chain~(${\ell > \abs{\hChainAfter}}$), those blocks are shifted to be pre-fork~(${a_d \gets a_d + \ell}$, ${a_c \gets a_c - \ell}$), and the public chain is also shifted to be pre-fork~(${h_d \gets h_d + \abs{\hChainAfter}}$, ${\hChainAfter \gets \emptyChain}$).
If ${\hap = \textsc{false}}$ (honest miners' blocks that are not part of the canonical chain cannot become acceptable) or if the number of pre-fork blocks (after the shift) is more than~$\forkSensitivity$, then the selfish miner's pre-fork blocks are uncontested, causing~$\hap \gets \textsc{false}$ and yielding a reward and difficulty contribution of~$a_d$.
An additional reward based on the transactions in the revealed~$\ell$ blocks is also yielded, and whale transactions that were in the revealed blocks are decremented from~$\pool$~(${\pool \gets \pool - \transactions{\hChainAfter[:\ell]}}$).
The values of~$c$ and \fork are reset to 0 and \forkIrrelevant, respectively.

If the selfish miner reveals a chain equal in length to the public chain~(${\ell = \abs{\hChainAfter}}$), one of the following cases occurs.
With the canonical ledger function, if the tie-breaking rule is worst-case, then the same as the previous case occurs.
If the tie-breaking rule is different, \fork is updated to \forkActive.

With the MAD ledger function, if the tie-breaking rule is worst-case, then the selfish miner's chain becomes the canonical chain~(${c \gets \abs{\hChainAfter}}$) and the blocks in the public chain are destructed~(${\transactions{\hChainAfter} \gets 0}$).
If the tie-breaking rule is different, if there are transactions in the public chain~(${\transactions{\hChainAfter} > 0}$), then the blocks in the public chain are destructed~(${\transactions{\hChainAfter} \gets 0}$) and nothing else happens (the blocks are not actually revealed); otherwise, \fork is updated to \forkActive.

\paragraph{Mine}

The selfish miner attempts mining a block pointing to the last block in their chain.
Honest miner attempt mining a block pointing to the last block in the public chain, and if \fork is \forkActive, also to the block in the selfish miner's chain with the same height.

The selfish miner succeeds~(${a_c \gets a_c + 1}$) with probability~$\alpha$.
In that case, \fork is updated to \forkIrrelevant.

If \fork is not \forkActive, the honest miners create a new block pointing to the last block in the public chain and include a transaction if there is one available~(${\hChainAfter \gets \hChainAfter\increment}$).
They succeed with probability~${1 - \alpha}$.

Otherwise, if \fork is \forkActive, there are two possible transitions.
First, with probability~${(1 -\gamma)(1 - \alpha)}$, an honest miner creates a block pointing to the block in the public chain first, indicating they prefer it to be in the canonical chain~(${\hChainAfter \gets \hChainAfter\increment}$).

With probability~${\gamma (1 - \alpha)}$, an honest miner creates a block pointing to the block in the selfish miner's block first.
The selfish miner's first~$\abs{\hChainAfter}$ blocks enter the canonical chain.
If the ledger function is the canonical one, the state is updated in the exact same manner as if the miner revealed a strictly longer chain with length~$\abs{\hChainAfter}$ (as in the Reveal~$\ell$ action).
If the ledger function is the MAD one, then only~$c$ is updated to~$\abs{\hChainAfter}$.

The pool is updated similarly to the full model, resulting in an identical formula for the utility of honest mining.

%%%%%%%%%%%%%%%%%%%%%%%%%%%%%%%%%%%%%%%%%%%%%%%%%%%%%%%%%%%%%%%%%%%%%%%%%%%%%%%%%%%%%%%%%%%%%%%%%%%%%%%%%%%%%%%%%%%%%%%%%%%%%%%%%%%%%%%%%%%%%%%%%%%%%%%%

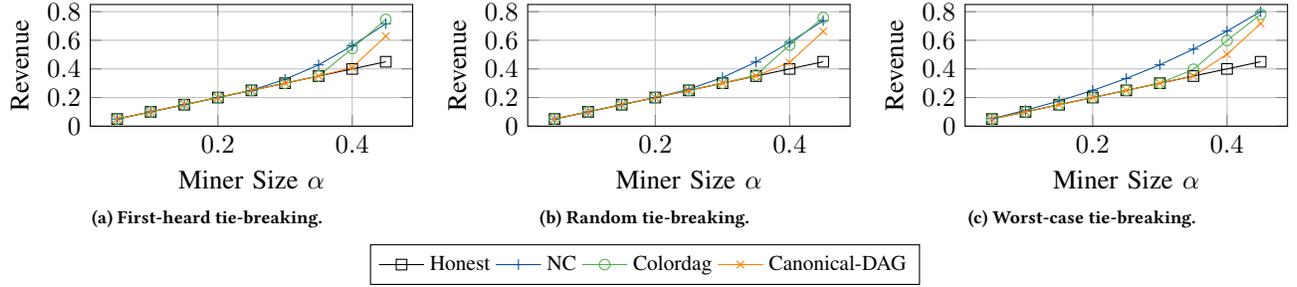
\begin{figure*}[t]
    \centering
    \subfloat[First-heard tie-breaking.]{\begin{tikzpicture}
            \pgfplotsset{height=0.18\textwidth,width=0.33\textwidth}
            \begin{axis}[
                xlabel={Miner Size~$\alpha$},
                ylabel={Revenue},
                legend pos=north west,
                addplot/.style={draw opacity=0.8},
                ymin=0,
                ymax=0.85,
                legend columns=4,
                legend to name=revenue-dam-legend
            ]
            \addplot table [x=alpha, y=Honest, only if all={tie_break_mode/random; model/bitcoin_fee; difficulty_source/uncontested}] {data/exact_results_no_fees.csv};
            \addlegendentry{Honest}

            \addplot table [x=alpha, y=ARR Revenue, only if all={tie_break_mode/first_heard; model/bitcoin_fee; difficulty_source/uncontested}] {data/exact_results_no_fees.csv};
            \addlegendentry{NC}

            \addplot table [x=alpha, y=ARR Revenue, only if all={tie_break_mode/first_heard; model/simplified_colordag; difficulty_source/uncontested}] {data/exact_results_no_fees.csv};
            \addlegendentry{Colordag}

            \addplot table [x=alpha, y=ARR Revenue, only if all={tie_break_mode/first_heard; model/simplified_colordag; difficulty_source/main}] {data/exact_results_no_fees.csv};
            \addlegendentry{Canonical-DAG}
            \end{axis}
        \end{tikzpicture}%
        \label{fig:revenue-dam:first-heard}}
    \hfil
    \subfloat[Random tie-breaking.]{\begin{tikzpicture}
            \pgfplotsset{height=0.18\textwidth,width=0.33\textwidth}
            \begin{axis}[
                xlabel={Miner Size~$\alpha$},
                ylabel={Revenue},
                addplot/.style={draw opacity=0.8},
                ymin=0,
                ymax=0.85,
            ]
            \addplot table [x=alpha, y=Honest, only if all={tie_break_mode/random; model/bitcoin_fee; difficulty_source/uncontested}] {data/exact_results_no_fees.csv};

            \addplot table [x=alpha, y=ARR Revenue, only if all={tie_break_mode/random; model/bitcoin_fee; difficulty_source/uncontested}] {data/exact_results_no_fees.csv};

            \addplot table [x=alpha, y=ARR Revenue, only if all={tie_break_mode/random; model/simplified_colordag; difficulty_source/uncontested}] {data/exact_results_no_fees.csv};

            \addplot table [x=alpha, y=ARR Revenue, only if all={tie_break_mode/random; model/simplified_colordag; difficulty_source/main}] {data/exact_results_no_fees.csv};
            \end{axis}
        \end{tikzpicture}%
        \label{fig:revenue-dam:random}}
    \hfil
    \subfloat[Worst-case tie-breaking.]{\begin{tikzpicture}
            \pgfplotsset{height=0.18\textwidth,width=0.33\textwidth}
            \begin{axis}[
                xlabel={Miner Size~$\alpha$},
                ylabel={Revenue},
                addplot/.style={draw opacity=0.8},
                ymin=0,
                ymax=0.85,
            ]
            \addplot table [x=alpha, y=Honest, only if all={tie_break_mode/random; model/bitcoin_fee; difficulty_source/uncontested}] {data/exact_results_no_fees.csv};

            \addplot table [x=alpha, y=ARR Revenue, only if all={tie_break_mode/attacker; model/bitcoin_fee; difficulty_source/uncontested}] {data/exact_results_no_fees.csv};

            \addplot table [x=alpha, y=ARR Revenue, only if all={tie_break_mode/attacker; model/simplified_colordag; difficulty_source/uncontested}] {data/exact_results_no_fees.csv};

            \addplot table [x=alpha, y=ARR Revenue, only if all={tie_break_mode/attacker; model/simplified_colordag; difficulty_source/main}] {data/exact_results_no_fees.csv};
            \end{axis}
        \end{tikzpicture}%
        \label{fig:revenue-dam:attacker}}
    \\
    \vspace*{0.5em}
    \pgfplotslegendfromname{revenue-dam-legend}
    
    \caption{Selfish mining revenue in NC, Colordag, and MAD-DAG.}
    \label{fig:revenue-dam}
    \Description{Selfish mining revenue in NC, Colordag, and MAD-DAG.}
\end{figure*}

\subsection{Solving the MDP}
\label{sec:analyzing-selfish-mining:solving-the-mdp}

%%%%%%%%%%%%%%%%%%%%%%%%%%%%%%%%%%%%%%%%%%%%%%%%%%%%%%%%%%%%%%%%%%%%%%%%%%%%%%%%%%%%%%%%%%%%%%%%%%%%%%%%%%%%%%%%%%%%%%%%%%%%%%%%%%%%%%%%%%%%%%%%%%%%%%%%

To solve all MDPs, we use PTO~\cite{zur2020efficient} to transform the MDP to a stochastic shortest path problem with expected horizon~$10^5$, and then solve the resultant MDP with policy iteration~\cite{bertsekas2012dynamic} with a precision of $10^{-5}$.
Using policy iteration, a dynamic programming algorithm, guarantees convergence to a near-optimal policy, allowing us to accurately calculate the security threshold.

Note that this is the first work to analyze NC with varying rewards with dynamic programming, and to obtain an accurate security threshold for NC.
This was made possible by several practical improvements over previous work.

% Trim state space
First, we trim the state space by removing infeasible states.
For example, a state where the public chain has more transactions than \pool is infeasible.
We also group states that are equivalent.
For example, the value of \fork does not matter if the selfish miner has fewer blocks than the public chain.

% Iterative linear solver for less memory (bicgstab)
Second, we use a more memory-efficient linear solver in each step of the policy iteration.
Policy iteration is an iterative algorithm where in each step, given the current policy, we compute the future utility of starting from each state by solving a linear system of equations, and then update the policy to pick the action that maximizes the future utility.
Instead of using a direct sparse linear solver as in previous work~\cite{zur2020efficient}, we use an iterative solver, Biconjugate gradient method~\cite{fletcher2006conjugate}, which requires less memory, a critical difference when the state space is large.
This solver sometimes does not converge to a solution.
When that happens we use another variant, Bi-conjugate gradient stabilized method~\cite{van1992bi}.

% Restrict \maxPool
Third, in most cases, we restrict \maxPool to~2.
This way, the number of possible chains of a certain length grows quadratically instead of exponentially.
We later verify that this restriction has little effect on the results for the cases consider ($\delta = 0.01$), but intuitively, it is simply highly unlikely that~3 whale transactions would be available at the same time.

Nevertheless, memory requirements remain significant;
in all our experiments, we utilize for a server with a large amount of RAM~(2 terabytes).

To find the security threshold, we simply perform a binary search to find the minimum~$\alpha$ required for profitable selfish mining.
In each step, we numerically calculate the utility of the output policy and compare it with the revenue of an honest miner with the current~$\alpha$ value.

%%%%%%%%%%%%%%%%%%%%%%%%%%%%%%%%%%%%%%%%%%%%%%%%%%%%%%%%%%%%%%%%%%%%%%%%%%%%%%%%%%%%%%%%%%%%%%%%%%%%%%%%%%%%%%%%%%%%%%%%%%%%%%%%%%%%%%%%%%%%%%%%%%%%%%%%%%%%%%%%%%%%%%%%%%%%%%%%%%%%%%%%%%%%%%%%%%%%%%%%%%%%%%%%%%%%%%%%%%%%%%%%%%%%%%%%%%%%%%%%%%%%%%%%%%%%%%%%%%%%%%%%%%%%%%%%%%%%%%%%%%%%%%%%%%%%%%%%%%%%%%%%%%%%%%%%%%%%%%%%%%%%%%%%%%%%%%%%%%%%%%%%%%%%%%%%%%%%%%%%%%%%%%%%%%%%%%%%%%%%%%%%%%%%%%%%%%%%%%%%%%%%%%%%%%%%%%%%%%%%%%%%%%%%%%%%%%%%%%%%%%%%%%%%%%%%%%%%

\section{Difficulty-Adjustment Mechanism}
\label{sec:difficulty-adjustment-mechanism}

%%%%%%%%%%%%%%%%%%%%%%%%%%%%%%%%%%%%%%%%%%%%%%%%%%%%%%%%%%%%%%%%%%%%%%%%%%%%%%%%%%%%%%%%%%%%%%%%%%%%%%%%%%%%%%%%%%%%%%%%%%%%%%%%%%%%%%%%%%%%%%%%%%%%%%%%%%%%%%%%%%%%%%%%%%%%%%%%%%%%%%%%%%%%%%%%%%%%%%%%%%%%%%%%%%%%%%%%%%%%%%%%%%%%%%%%%%%%%%%%%%%%%%%%%%%%%%%%%%%%%%%%%%%%%%%%%%%%%%%%%%%%%%%%%%%%%%%%%%%%%%%%%%%%%%%%%%%%%%%%%%%%%%%%%%%%%%%%%%%%%%%%%%%%%%%%%%%%%%%%%%%%%%%%%%%%%%%%%%%%%%%%%%%%%%%%%%%%%%%%%%%%%%%%%%%%%%%%%%%%%%%%%%%%%%%%%%%%%%%%%%%%%%%%%%%%%%%%

% Rest of difficulty-adjustment figures

\pgfplotstableread[col sep=comma]{data/bitcoin_no_fees.csv}\bitcoinTableNoFees
\pgfplotstablefindvalue[0]{\bitcoinTableNoFees}{tie_break_mode/first_heard}{Threshold}{\ThresholdBitcoinNoFeesFirstHeard}
\pgfplotstablefindvalue[0]{\bitcoinTableNoFees}{tie_break_mode/random}{Threshold}{\ThresholdBitcoinNoFeesRandom}
\pgfplotstablefindvalue[0]{\bitcoinTableNoFees}{tie_break_mode/attacker}{Threshold}{\ThresholdBitcoinNoFeesAttacker}

\begin{figure*}[t]
    \centering
    \subfloat[First-heard tie-breaking.]{%
        \begin{tikzpicture}
            \pgfplotsset{height=0.18\textwidth,width=0.33\textwidth}
            \begin{axis}[
                xlabel={Fork Sensitivity~$\forkSensitivity$},
                ylabel={Security Threshold},
                legend pos=south east,
                addplot/.style={draw opacity=0.8},
                ymin=-0.05,
                ymax=0.45,
                cycle list shift=1,
                legend columns=3,
                legend to name=fork-sensitivity-dam-legend
            ]
            \addplot table [x=acceptable_path_param, y expr=\ThresholdBitcoinNoFeesFirstHeard, only if all={tie_break_mode/first_heard; difficulty_source/uncontested; max_fork/20}] {data/simplified_colordag_model_no_fees.csv};
            \addlegendentry{NC}

            \addplot table [x=acceptable_path_param, y=Threshold, only if all={tie_break_mode/first_heard; difficulty_source/uncontested; max_fork/20}] {data/simplified_colordag_model_no_fees.csv};
            \addlegendentry{Colordag}

            \addplot table [x=acceptable_path_param, y=Threshold, only if all={tie_break_mode/first_heard; difficulty_source/main; max_fork/20}] {data/simplified_colordag_model_no_fees.csv};
            \addlegendentry{Canonical-DAG}
            \end{axis}
        \end{tikzpicture}%
        \label{fig:fork-sensitivity-dam:first-heard}}
    \hfil
    \subfloat[Random tie-breaking.]{%
        \begin{tikzpicture}
            \pgfplotsset{height=0.18\textwidth,width=0.33\textwidth}
            \begin{axis}[
                xlabel={Fork Sensitivity~$\forkSensitivity$},
                ylabel={Security Threshold},
                addplot/.style={draw opacity=0.8},
                ymin=-0.05,
                ymax=0.45,
                cycle list shift=1,
            ]
            \addplot table [x=acceptable_path_param, y expr=\ThresholdBitcoinNoFeesRandom, only if all={tie_break_mode/random; difficulty_source/uncontested; max_fork/20}] {data/simplified_colordag_model_no_fees.csv};

            \addplot table [x=acceptable_path_param, y=Threshold, only if all={tie_break_mode/random; difficulty_source/uncontested; max_fork/20}] {data/simplified_colordag_model_no_fees.csv};

            \addplot table [x=acceptable_path_param, y=Threshold, only if all={tie_break_mode/random; difficulty_source/main; max_fork/20}] {data/simplified_colordag_model_no_fees.csv};
            \end{axis}
        \end{tikzpicture}%
        \label{fig:fork-sensitivity-dam:random}}
    \hfil
    \subfloat[Worst-case tie-breaking.]{%
        \begin{tikzpicture}
            \pgfplotsset{height=0.18\textwidth,width=0.33\textwidth}
            \begin{axis}[
                xlabel={Fork Sensitivity~$\forkSensitivity$},
                ylabel={Security Threshold},
                addplot/.style={draw opacity=0.8},
                ymin=-0.05,
                ymax=0.45,
                cycle list shift=1,
            ]
            \addplot table [x=acceptable_path_param, y expr=\ThresholdBitcoinNoFeesAttacker, only if all={tie_break_mode/attacker; difficulty_source/uncontested; max_fork/20}] {data/simplified_colordag_model_no_fees.csv};

            \addplot table [x=acceptable_path_param, y=Threshold, only if all={tie_break_mode/attacker; difficulty_source/uncontested; max_fork/20}] {data/simplified_colordag_model_no_fees.csv};

            \addplot table [x=acceptable_path_param, y=Threshold, only if all={tie_break_mode/attacker; difficulty_source/main; max_fork/20}] {data/simplified_colordag_model_no_fees.csv};
            \end{axis}
        \end{tikzpicture}%
        \label{fig:threshold-dam:attacker}}
    \\
    \vspace*{0.5em}
    \pgfplotslegendfromname{fork-sensitivity-dam-legend}

    \caption{Security threshold of NC, Colordag, and MAD-DAG as a function of fork sensitivity.}
    \label{fig:threshold-dam}
    \Description{Security threshold of NC, Colordag, and MAD-DAG as a function of fork sensitivity.}
\end{figure*}
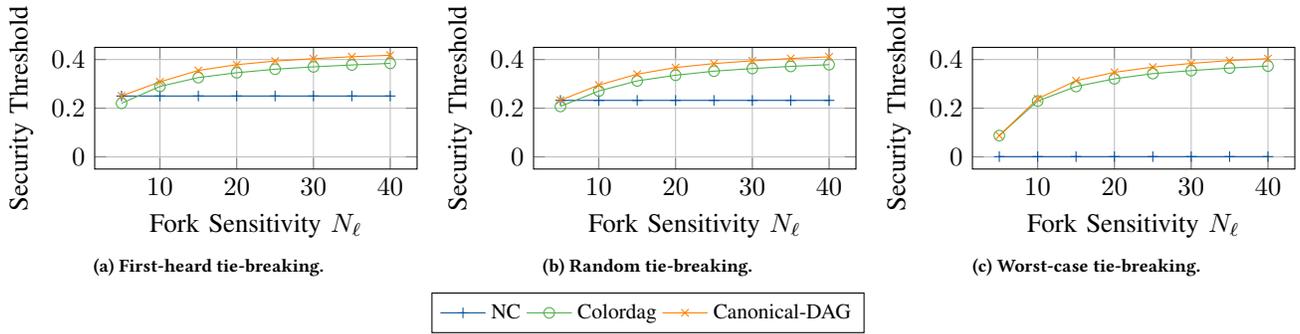

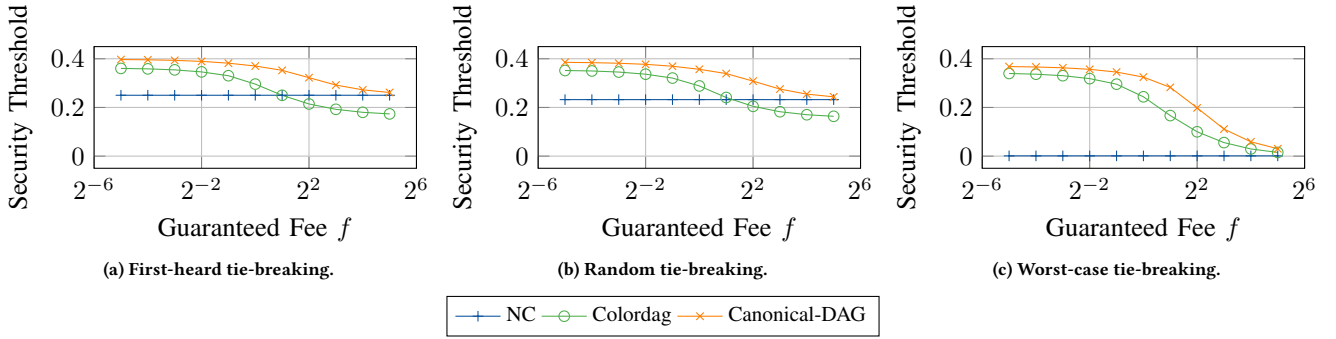
\begin{figure*}[t]
    \centering
    \subfloat[First-heard tie-breaking.]{%
        \begin{tikzpicture}
            \pgfplotsset{height=0.18\textwidth,width=0.33\textwidth}
            \begin{axis}[
                xlabel={Guaranteed Fee~$f$},
                ylabel={Security Threshold},
                addplot/.style={draw opacity=0.8},
                ymin=-0.05,
                ymax=0.45,
                xmode=log,
                log basis x=2,
                legend pos=south west,
                cycle list shift=1,
                legend columns=3,
                legend to name=guaranteed-fee-dam-legend
            ]

            \addplot table [x=guaranteed_fee, y expr=\ThresholdBitcoinNoFeesFirstHeard, only if all={tie_break_mode/first_heard; difficulty_source/uncontested}] {data/constant_fees_no_whales.csv};
            \addlegendentry{NC}

            \addplot table [x=guaranteed_fee, y=Threshold, only if all={tie_break_mode/first_heard; difficulty_source/uncontested}] {data/constant_fees_no_whales.csv};
            \addlegendentry{Colordag}

            \addplot table [x=guaranteed_fee, y=Threshold, only if all={tie_break_mode/first_heard; difficulty_source/main}] {data/constant_fees_no_whales.csv};
            \addlegendentry{Canonical-DAG}
            \end{axis}
        \end{tikzpicture}%
        \label{fig:guaranteed-fee-dam:first-heard}}
    \hfil
    \subfloat[Random tie-breaking.]{%
        \begin{tikzpicture}
            \pgfplotsset{height=0.18\textwidth,width=0.33\textwidth}
            \begin{axis}[
                xlabel={Guaranteed Fee~$f$},
                ylabel={Security Threshold},
                addplot/.style={draw opacity=0.8},
                ymin=-0.05,
                ymax=0.45,
                xmode=log,
                log basis x=2,            
                cycle list shift=1,
            ]
            \addplot table [x=guaranteed_fee, y expr=\ThresholdBitcoinNoFeesRandom, only if all={tie_break_mode/first_heard; difficulty_source/uncontested}] {data/constant_fees_no_whales.csv};

            \addplot table [x=guaranteed_fee, y=Threshold, only if all={tie_break_mode/random; difficulty_source/uncontested}] {data/constant_fees_no_whales.csv};

            \addplot table [x=guaranteed_fee, y=Threshold, only if all={tie_break_mode/random; difficulty_source/main}] {data/constant_fees_no_whales.csv};
            \end{axis}
        \end{tikzpicture}%
        \label{fig:guaranteed-fee-dam:random}}
    \hfil
    \subfloat[Worst-case tie-breaking.]{%
        \begin{tikzpicture}
            \pgfplotsset{height=0.18\textwidth,width=0.33\textwidth}
            \begin{axis}[
                xlabel={Guaranteed Fee~$f$},
                ylabel={Security Threshold},
                addplot/.style={draw opacity=0.8},
                ymin=-0.05,
                ymax=0.45,
                xmode=log,
                log basis x=2,            
                cycle list shift=1,
            ]
            \addplot table [x=guaranteed_fee, y expr=\ThresholdBitcoinNoFeesAttacker, only if all={tie_break_mode/attacker; difficulty_source/uncontested}] {data/constant_fees_no_whales.csv};

            \addplot table [x=guaranteed_fee, y=Threshold, only if all={tie_break_mode/attacker; difficulty_source/uncontested}] {data/constant_fees_no_whales.csv};

            \addplot table [x=guaranteed_fee, y=Threshold, only if all={tie_break_mode/attacker; difficulty_source/main}] {data/constant_fees_no_whales.csv};
            \end{axis}
        \end{tikzpicture}%
        \label{fig:guaranteed-fee-dam:attacker}}
    \\
    \vspace*{0.5em}
    \pgfplotslegendfromname{guaranteed-fee-dam-legend}

    \caption{Security threshold of NC, Colordag, and MAD-DAG as a function of guaranteed fee~$f$.}
    \label{fig:guaranteed-fee-dam}
    \Description{Security threshold of NC, Colordag, and MAD-DAG as a function of guaranteed fee~$f$.}
\end{figure*}

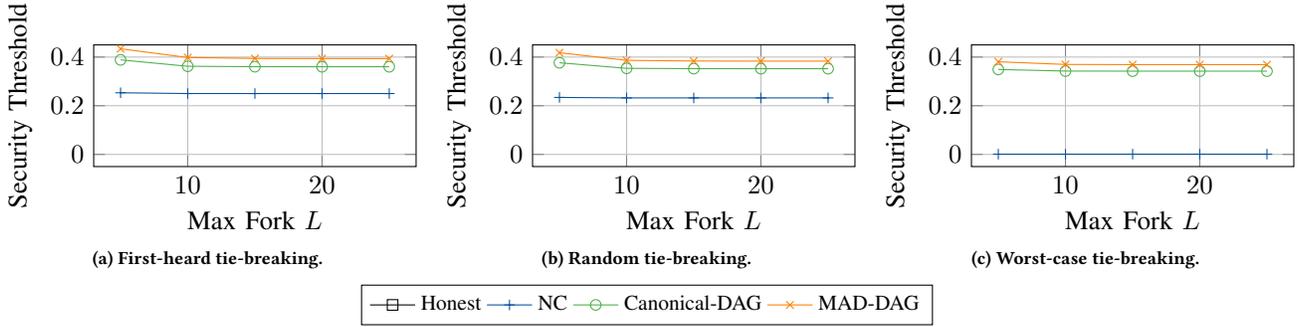
\begin{figure*}[t]
    \centering
    \subfloat[First-heard tie-breaking.]{%
        \begin{tikzpicture}
            \pgfplotsset{height=0.18\textwidth,width=0.33\textwidth}
            \begin{axis}[
                xlabel={Max Fork~$L$},
                ylabel={Security Threshold},
                legend pos=south east,
                addplot/.style={draw opacity=0.8},
                ymin=-0.05,
                ymax=0.45,
                cycle list shift=1,
                legend columns=3,
                legend to name=max-fork-dam-legend
            ]
            \addplot table [x=max_fork, y=Threshold, only if all={tie_break_mode/first_heard}] {data/max_fork_bitcoin_no_fees.csv};
            \addlegendentry{NC}

            \addplot table [x=max_fork, y=Threshold, only if all={tie_break_mode/first_heard; difficulty_source/uncontested}] {data/max_fork_no_fees.csv};
            \addlegendentry{Colordag}

            \addplot table [x=max_fork, y=Threshold, only if all={tie_break_mode/first_heard; difficulty_source/main}] {data/max_fork_no_fees.csv};
            \addlegendentry{Canonical-DAG}
            \end{axis}
        \end{tikzpicture}%
        \label{fig:max-fork-dam:first-heard}}
    \hfil
    \subfloat[Random tie-breaking.]{%
        \begin{tikzpicture}
            \pgfplotsset{height=0.18\textwidth,width=0.33\textwidth}
            \begin{axis}[
                xlabel={Max Fork~$L$},
                ylabel={Security Threshold},
                addplot/.style={draw opacity=0.8},
                ymin=-0.05,
                ymax=0.45,
                cycle list shift=1,
            ]
            \addplot table [x=max_fork, y=Threshold, only if all={tie_break_mode/random}] {data/max_fork_bitcoin_no_fees.csv};

            \addplot table [x=max_fork, y=Threshold, only if all={tie_break_mode/random; difficulty_source/uncontested}] {data/max_fork_no_fees.csv};

            \addplot table [x=max_fork, y=Threshold, only if all={tie_break_mode/random; difficulty_source/main}] {data/max_fork_no_fees.csv};
            \end{axis}
        \end{tikzpicture}%
        \label{fig:max-fork-dam:random}}
    \hfil
    \subfloat[Worst-case tie-breaking.]{%
        \begin{tikzpicture}
            \pgfplotsset{height=0.18\textwidth,width=0.33\textwidth}
            \begin{axis}[
                xlabel={Max Fork~$L$},
                ylabel={Security Threshold},
                addplot/.style={draw opacity=0.8},
                ymin=-0.05,
                ymax=0.45,
                cycle list shift=1,
            ]
            \addplot table [x=max_fork, y=Threshold, only if all={tie_break_mode/attacker}] {data/max_fork_bitcoin_no_fees.csv};

            \addplot table [x=max_fork, y=Threshold, only if all={tie_break_mode/attacker; difficulty_source/uncontested}] {data/max_fork_no_fees.csv};

            \addplot table [x=max_fork, y=Threshold, only if all={tie_break_mode/attacker; difficulty_source/main}] {data/max_fork_no_fees.csv};
            \end{axis}
        \end{tikzpicture}%
        \label{fig:max-fork-dam:attacker}}
    \\
    \vspace*{0.5em}
    \pgfplotslegendfromname{max-fork-dam-legend}

    \caption{Security threshold of NC, Colordag, and MAD-DAG as a function of maximum fork length~$L$.}
    \label{fig:max-fork-dam}
    \Description{Security threshold of NC, Colordag, and MAD-DAG as a function of maximum fork length~$L$.}
\end{figure*}

\begin{figure*}[t]
    \centering
    \subfloat[First-heard tie-breaking.]{%
        \begin{tikzpicture}
            \pgfplotsset{height=0.18\textwidth,width=0.33\textwidth}
            \begin{axis}[
                xlabel={Max Fork~$L$},
                ylabel={Security Threshold},
                legend pos=south east,
                addplot/.style={draw opacity=0.8},
                ymin=-0.05,
                ymax=0.55,
                cycle list shift=2,
                legend columns=4,
                legend to name=max-fork-dam-comparison-legend
            ]
            \addplot table [x=max_fork, y=Threshold, only if all={tie_break_mode/first_heard; difficulty_source/uncontested; ledger_function/longest; model/simplified_colordag}] {data/colordag_model_comparison_max_fork.csv};
            \addlegendentry{Colordag (Upper-bound)}

            \addplot table [x=max_fork, y=Threshold, only if all={tie_break_mode/first_heard; difficulty_source/main; ledger_function/longest; model/simplified_colordag}] {data/colordag_model_comparison_max_fork.csv};
            \addlegendentry{Canonical-DAG (Upper-bound)}
            
            \addplot table [x=max_fork, y=Threshold, only if all={tie_break_mode/first_heard; difficulty_source/uncontested; ledger_function/longest; model/chain_colordag}] {data/colordag_model_comparison_max_fork.csv};
            \addlegendentry{Colordag (Full)}

            \addplot table [x=max_fork, y=Threshold, only if all={tie_break_mode/first_heard; difficulty_source/main; ledger_function/longest; model/chain_colordag}] {data/colordag_model_comparison_max_fork.csv};
            \addlegendentry{Canonical-DAG (Full)}
            \end{axis}
        \end{tikzpicture}%
        \label{fig:max-fork-dam-comparison:first-heard}}
    \hfil
    \subfloat[Random tie-breaking.]{%
        \begin{tikzpicture}
            \pgfplotsset{height=0.18\textwidth,width=0.33\textwidth}
            \begin{axis}[
                xlabel={Max Fork~$L$},
                ylabel={Security Threshold},
                addplot/.style={draw opacity=0.8},
                ymin=-0.05,
                ymax=0.55,
                cycle list shift=2,
            ]
            \addplot table [x=max_fork, y=Threshold, only if all={tie_break_mode/random; difficulty_source/uncontested; ledger_function/longest; model/simplified_colordag}] {data/colordag_model_comparison_max_fork.csv};

            \addplot table [x=max_fork, y=Threshold, only if all={tie_break_mode/random; difficulty_source/main; ledger_function/longest; model/simplified_colordag}] {data/colordag_model_comparison_max_fork.csv};

            \addplot table [x=max_fork, y=Threshold, only if all={tie_break_mode/random; difficulty_source/uncontested; ledger_function/longest; model/chain_colordag}] {data/colordag_model_comparison_max_fork.csv};

            \addplot table [x=max_fork, y=Threshold, only if all={tie_break_mode/random; difficulty_source/main; ledger_function/longest; model/chain_colordag}] {data/colordag_model_comparison_max_fork.csv};
            \end{axis}
        \end{tikzpicture}%
        \label{fig:max-fork-dam-comparison:random}}
    \hfil
    \subfloat[Worst-case tie-breaking.]{%
        \begin{tikzpicture}
            \pgfplotsset{height=0.18\textwidth,width=0.33\textwidth}
            \begin{axis}[
                xlabel={Max Fork~$L$},
                ylabel={Security Threshold},
                addplot/.style={draw opacity=0.8},
                ymin=-0.05,
                ymax=0.55,
                cycle list shift=2,
            ]
            \addplot table [x=max_fork, y=Threshold, only if all={tie_break_mode/attacker; difficulty_source/uncontested; ledger_function/longest; model/simplified_colordag}] {data/colordag_model_comparison_max_fork.csv};

            \addplot table [x=max_fork, y=Threshold, only if all={tie_break_mode/attacker; difficulty_source/main; ledger_function/longest; model/simplified_colordag}] {data/colordag_model_comparison_max_fork.csv};

            \addplot table [x=max_fork, y=Threshold, only if all={tie_break_mode/attacker; difficulty_source/uncontested; ledger_function/longest; model/chain_colordag}] {data/colordag_model_comparison_max_fork.csv};

            \addplot table [x=max_fork, y=Threshold, only if all={tie_break_mode/attacker; difficulty_source/main; ledger_function/longest; model/chain_colordag}] {data/colordag_model_comparison_max_fork.csv};
            \end{axis}
        \end{tikzpicture}%
        \label{fig:max-fork-dam-comparison:attacker}}
    \\
    \vspace*{0.5em}
    \pgfplotslegendfromname{max-fork-dam-comparison-legend}

    \caption{Comparison of security threshold,  Colordag, and MAD-DAG in the simplified model and the full model as a function of maximum fork length~$L$.}
    \label{fig:max-fork-dam-comparison}
    \Description{Comparison of security threshold,  Colordag, and MAD-DAG in the simplified model and the full model as a function of maximum fork length~$L$.}
\end{figure*}
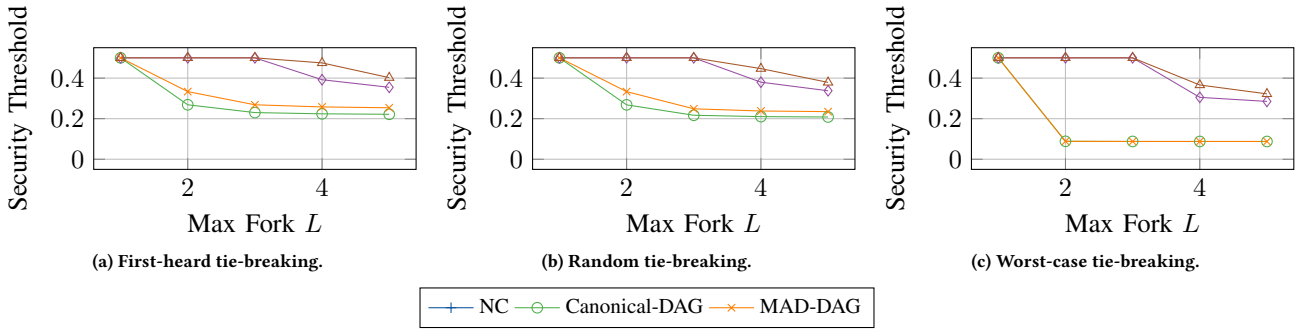

% Ledger function figures

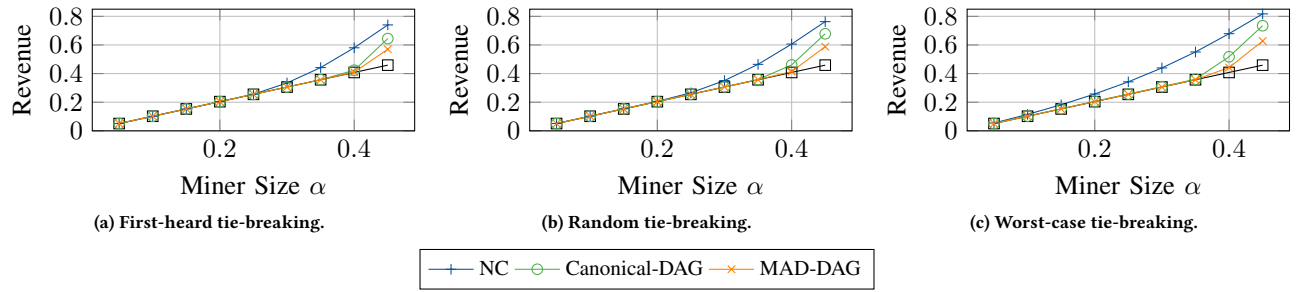
\begin{figure*}[t]
    \centering
    \subfloat[First-heard tie-breaking.]{%
        \begin{tikzpicture}
            \pgfplotsset{height=0.18\textwidth,width=0.33\textwidth}
            \begin{axis}[
                xlabel={Miner Size~$\alpha$},
                ylabel={Revenue},
                legend pos=north west,
                addplot/.style={draw opacity=0.8},
                ymin=0,
                ymax=0.85,
                legend columns=4,
                legend to name=revenue-lf-legend
            ]
            \addplot table [x=alpha, y=Honest, only if all={tie_break_mode/random; model/bitcoin_fee; ledger_function/longest}] {data/exact_results_with_fees.csv};
            \addlegendentry{Honest}

            \addplot table [x=alpha, y=ARR Revenue, only if all={tie_break_mode/first_heard; model/bitcoin_fee;ledger_function/longest}] {data/exact_results_with_fees.csv};
            \addlegendentry{NC}

            \addplot table [x=alpha, y=ARR Revenue, only if all={tie_break_mode/first_heard; model/simplified_colordag; ledger_function/longest}] {data/exact_results_with_fees.csv};
            \addlegendentry{Canonical-DAG}

            \addplot table [x=alpha, y=ARR Revenue, only if all={tie_break_mode/first_heard; model/simplified_colordag; ledger_function/mad}] {data/exact_results_with_fees.csv};
            \addlegendentry{MAD-DAG}
            \end{axis}
        \end{tikzpicture}%
        \label{fig:revenue-lf:first-heard}}
    \hfil
    \subfloat[Random tie-breaking.]{%
        \begin{tikzpicture}
            \pgfplotsset{height=0.18\textwidth,width=0.33\textwidth}
            \begin{axis}[
                xlabel={Miner Size~$\alpha$},
                ylabel={Revenue},
                addplot/.style={draw opacity=0.8},
                ymin=0,
                ymax=0.85,
            ]
            \addplot table [x=alpha, y=Honest, only if all={tie_break_mode/random; model/bitcoin_fee; ledger_function/longest}] {data/exact_results_with_fees.csv};

            \addplot table [x=alpha, y=ARR Revenue, only if all={tie_break_mode/random; model/bitcoin_fee; ledger_function/longest}] {data/exact_results_with_fees.csv};

            \addplot table [x=alpha, y=ARR Revenue, only if all={tie_break_mode/random; model/simplified_colordag; ledger_function/longest}] {data/exact_results_with_fees.csv};

            \addplot table [x=alpha, y=ARR Revenue, only if all={tie_break_mode/random; model/simplified_colordag; ledger_function/mad}] {data/exact_results_with_fees.csv};
            \end{axis}
        \end{tikzpicture}%
        \label{fig:revenue-lf:random}}
    \hfil
    \subfloat[Worst-case tie-breaking.]{%
        \begin{tikzpicture}
            \pgfplotsset{height=0.18\textwidth,width=0.33\textwidth}
            \begin{axis}[
                xlabel={Miner Size~$\alpha$},
                ylabel={Revenue},
                addplot/.style={draw opacity=0.8},
                ymin=0,
                ymax=0.85,
            ]
            \addplot table [x=alpha, y=Honest, only if all={tie_break_mode/random; model/bitcoin_fee; ledger_function/longest}] {data/exact_results_with_fees.csv};

            \addplot table [x=alpha, y=ARR Revenue, only if all={tie_break_mode/attacker; model/bitcoin_fee; ledger_function/longest}] {data/exact_results_with_fees.csv};

            \addplot table [x=alpha, y=ARR Revenue, only if all={tie_break_mode/attacker; model/simplified_colordag; ledger_function/longest}] {data/exact_results_with_fees.csv};

            \addplot table [x=alpha, y=ARR Revenue, only if all={tie_break_mode/attacker; model/simplified_colordag; ledger_function/mad}] {data/exact_results_with_fees.csv};
            \end{axis}
        \end{tikzpicture}%
        \label{fig:revenue-lf:attacker}}
    \\
    \vspace*{0.5em}
    \pgfplotslegendfromname{revenue-lf-legend}

    \caption{Selfish mining revenue in NC, Colordag, and MAD-DAG with whale transactions.}
    \label{fig:revenue-lf}
    \Description{Selfish mining revenue in NC, Colordag, and MAD-DAG with whale transactions.}
\end{figure*}

\pgfplotstableread[col sep=comma]{data/bitcoin_with_fees.csv}\bitcoinTableWithFees
\pgfplotstablefindvalue[0]{\bitcoinTableWithFees}{tie_break_mode/first_heard; fee/2}{Threshold}{\ThresholdBitcoinWithFeesFirstHeard}
\pgfplotstablefindvalue[0]{\bitcoinTableWithFees}{tie_break_mode/random; fee/2}{Threshold}{\ThresholdBitcoinWithFeesRandom}
\pgfplotstablefindvalue[0]{\bitcoinTableWithFees}{tie_break_mode/attacker; fee/2}{Threshold}{\ThresholdBitcoinWithFeesAttacker}

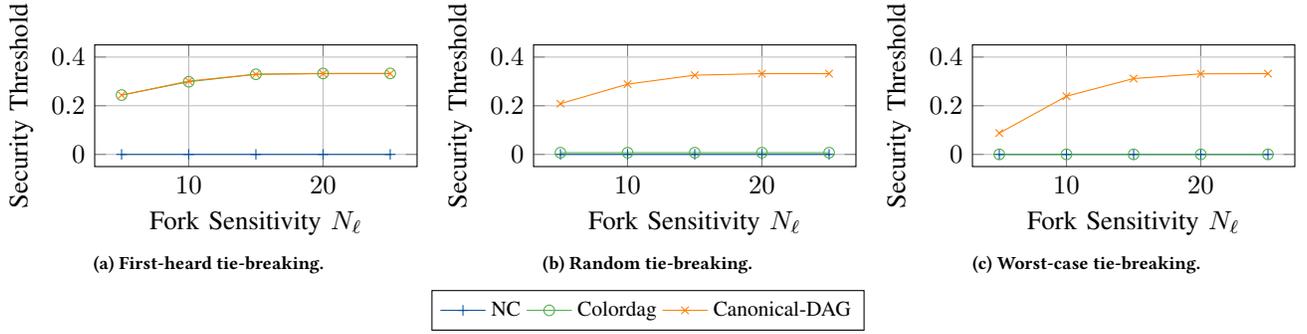
\begin{figure*}[t]
    \centering
    \subfloat[First-heard tie-breaking.]{%
        \begin{tikzpicture}
            \pgfplotsset{height=0.18\textwidth,width=0.33\textwidth}
            \begin{axis}[
                xlabel={Fork Sensitivity~$\forkSensitivity$},
                ylabel={Security Threshold},
                legend pos=south east,
                addplot/.style={draw opacity=0.8},
                ymin=-0.05,
                ymax=0.45,
                cycle list shift=1,
                legend columns=3,
                legend to name=fork-sensitivity-lf-legend
            ]
            \addplot table [x=acceptable_path_param, y expr=\ThresholdBitcoinWithFeesFirstHeard, only if all={tie_break_mode/first_heard; ledger_function/longest}] {data/simplified_colordag_model_whales_nl.csv};
            \addlegendentry{NC}

            \addplot table [x=acceptable_path_param, y=Threshold, only if all={tie_break_mode/first_heard; ledger_function/longest}] {data/simplified_colordag_model_whales_nl.csv};
            \addlegendentry{Canonical-DAG}

            \addplot table [x=acceptable_path_param, y=Threshold, only if all={tie_break_mode/first_heard; ledger_function/mad}] {data/simplified_colordag_model_whales_nl.csv};
            \addlegendentry{MAD-DAG}
            \end{axis}
        \end{tikzpicture}%
        \label{fig:fork-sensitivity-lf:first-heard}}
    \hfil
    \subfloat[Random tie-breaking.]{%
        \begin{tikzpicture}
            \pgfplotsset{height=0.18\textwidth,width=0.33\textwidth}
            \begin{axis}[
                xlabel={Fork Sensitivity~$\forkSensitivity$},
                ylabel={Security Threshold},
                addplot/.style={draw opacity=0.8},
                ymin=-0.05,
                ymax=0.45,
                cycle list shift=1,
            ]
            \addplot table [x=acceptable_path_param, y expr=\ThresholdBitcoinWithFeesRandom, only if all={tie_break_mode/random; ledger_function/longest}] {data/simplified_colordag_model_whales_nl.csv};

            \addplot table [x=acceptable_path_param, y=Threshold, only if all={tie_break_mode/random; ledger_function/longest}] {data/simplified_colordag_model_whales_nl.csv};

            \addplot table [x=acceptable_path_param, y=Threshold, only if all={tie_break_mode/random; ledger_function/mad}] {data/simplified_colordag_model_whales_nl.csv};
            \end{axis}
        \end{tikzpicture}%
        \label{fig:fork-sensitivity-lf:random}}
    \hfil
    \subfloat[Worst-case tie-breaking.]{%
        \begin{tikzpicture}
            \pgfplotsset{height=0.18\textwidth,width=0.33\textwidth}
            \begin{axis}[
                xlabel={Fork Sensitivity~$\forkSensitivity$},
                ylabel={Security Threshold},
                addplot/.style={draw opacity=0.8},
                ymin=-0.05,
                ymax=0.45,
                cycle list shift=1,
            ]
            \addplot table [x=acceptable_path_param, y expr=\ThresholdBitcoinWithFeesAttacker, only if all={tie_break_mode/attacker; ledger_function/longest}] {data/simplified_colordag_model_whales_nl.csv};

            \addplot table [x=acceptable_path_param, y=Threshold, only if all={tie_break_mode/attacker; ledger_function/longest}] {data/simplified_colordag_model_whales_nl.csv};

            \addplot table [x=acceptable_path_param, y=Threshold, only if all={tie_break_mode/attacker; ledger_function/mad}] {data/simplified_colordag_model_whales_nl.csv};
            \end{axis}
        \end{tikzpicture}%
        \label{fig:fork-sensitivity-lf:attacker}}
    \\
    \vspace*{0.5em}
    \pgfplotslegendfromname{fork-sensitivity-lf-legend}

    \caption{Security threshold of NC, Canonical-DAG, and MAD-DAG as a function of fork sensitivity.}
    \label{fig:fork-sensitivity-lf}
    \Description{Security threshold of NC, Canonical-DAG, and MAD-DAG as a function of fork sensitivity.}
\end{figure*}

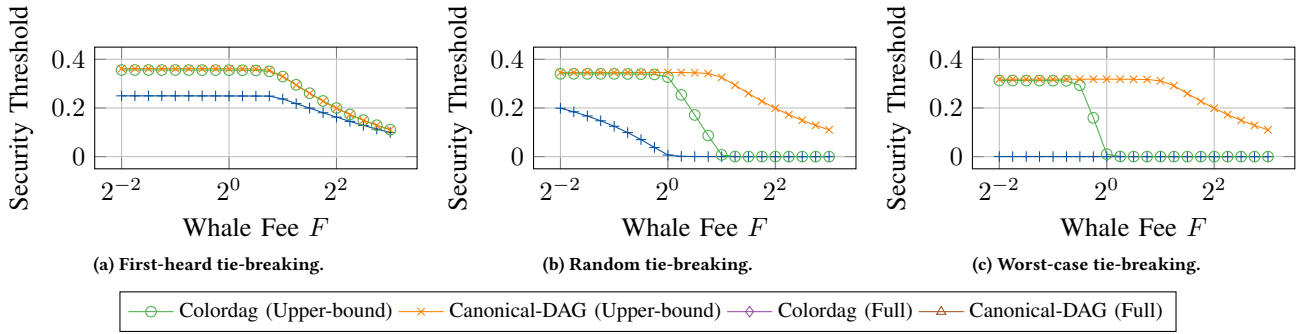
\begin{figure*}[t]
    \centering
    \subfloat[First-heard tie-breaking.]{%
        \begin{tikzpicture}
            \pgfplotsset{height=0.18\textwidth,width=0.33\textwidth}
            \begin{axis}[
                xlabel={Whale Fee~$F$},
                ylabel={Security Threshold},
                addplot/.style={draw opacity=0.8},
                ymin=-0.05,
                ymax=0.45,
                xmode=log,
                log basis x=2,
                legend pos=south west,
                cycle list shift=1,
                legend columns=3,
                legend to name=whale-fee-lf-legend
            ]

            \addplot table [x=fee, y=Threshold, only if all={tie_break_mode/first_heard}] {data/bitcoin_with_fees.csv};
            \addlegendentry{NC}

            \addplot table [x=fee, y=Threshold, only if all={tie_break_mode/first_heard; ledger_function/longest}] {data/simplified_colordag_model.csv};
            \addlegendentry{Canonical-DAG}

            \addplot table [x=fee, y=Threshold, only if all={tie_break_mode/first_heard; ledger_function/mad}] {data/simplified_colordag_model.csv};
            \addlegendentry{MAD-DAG}
            \end{axis}
        \end{tikzpicture}%
        \label{fig:whale-fee-lf:first-heard}}
    \hfil
    \subfloat[Random tie-breaking.]{%
        \begin{tikzpicture}
            \pgfplotsset{height=0.18\textwidth,width=0.33\textwidth}
            \begin{axis}[
                xlabel={Whale Fee~$F$},
                ylabel={Security Threshold},
                addplot/.style={draw opacity=0.8},
                ymin=-0.05,
                ymax=0.45,
                xmode=log,
                log basis x=2,            
                cycle list shift=1,
            ]
            \addplot table [x=fee, y=Threshold, only if all={tie_break_mode/random}] {data/bitcoin_with_fees.csv};

            \addplot table [x=fee, y=Threshold, only if all={tie_break_mode/random; ledger_function/longest}] {data/simplified_colordag_model.csv};

            \addplot table [x=fee, y=Threshold, only if all={tie_break_mode/random; ledger_function/mad}] {data/simplified_colordag_model.csv};
            \end{axis}
        \end{tikzpicture}%
        \label{fig:whale-fee-lf:random}}
    \hfil
    \subfloat[Worst-case tie-breaking.]{%
        \begin{tikzpicture}
            \pgfplotsset{height=0.18\textwidth,width=0.33\textwidth}
            \begin{axis}[
                xlabel={Whale Fee~$F$},
                ylabel={Security Threshold},
                addplot/.style={draw opacity=0.8},
                ymin=-0.05,
                ymax=0.45,
                xmode=log,
                log basis x=2,            
                cycle list shift=1,
            ]
            \addplot table [x=fee, y=Threshold, only if all={tie_break_mode/attacker}] {data/bitcoin_with_fees.csv};

            \addplot table [x=fee, y=Threshold, only if all={tie_break_mode/attacker; ledger_function/longest}] {data/simplified_colordag_model.csv};

            \addplot table [x=fee, y=Threshold, only if all={tie_break_mode/attacker; ledger_function/mad}] {data/simplified_colordag_model.csv};
            \end{axis}
        \end{tikzpicture}%
        \label{fig:whale-fee-lf:attacker}}
    \\
    \vspace*{0.5em}
    \pgfplotslegendfromname{whale-fee-lf-legend}

    \caption{Security threshold of NC, Canonical-DAG, and MAD-DAG as a function of whale fee~$F$.}
    \label{fig:whale-fee-lf}
    \Description{Security threshold of NC, Canonical-DAG, and MAD-DAG as a function of whale fee~$F$.}
\end{figure*}

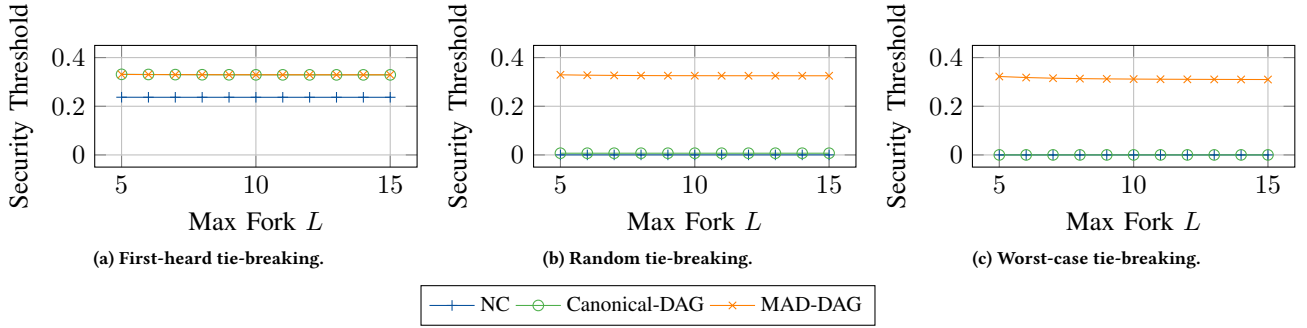
\begin{figure*}[t]
    \centering
    \subfloat[First-heard tie-breaking.]{%
        \begin{tikzpicture}
            \pgfplotsset{height=0.18\textwidth,width=0.33\textwidth}
            \begin{axis}[
                xlabel={Max Fork~$L$},
                ylabel={Security Threshold},
                legend pos=south east,
                addplot/.style={draw opacity=0.8},
                ymin=-0.05,
                ymax=0.45,
                cycle list shift=1,
                legend columns=3,
                legend to name=max-fork-lf-legend
            ]
            \addplot table [x=max_fork, y=Threshold, only if all={tie_break_mode/first_heard}] {data/max_fork_bitcoin_whales.csv};
            \addlegendentry{NC}

            \addplot table [x=max_fork, y=Threshold, only if all={tie_break_mode/first_heard; ledger_function/longest}] {data/max_fork_whales.csv};
            \addlegendentry{Canonical-DAG}

            \addplot table [x=max_fork, y=Threshold, only if all={tie_break_mode/first_heard; ledger_function/mad}] {data/max_fork_whales.csv};
            \addlegendentry{MAD-DAG}
            \end{axis}
        \end{tikzpicture}%
        \label{fig:max-fork-lf:first-heard}}
    \hfil
    \subfloat[Random tie-breaking.]{%
        \begin{tikzpicture}
            \pgfplotsset{height=0.18\textwidth,width=0.33\textwidth}
            \begin{axis}[
                xlabel={Max Fork~$L$},
                ylabel={Security Threshold},
                addplot/.style={draw opacity=0.8},
                ymin=-0.05,
                ymax=0.45,
                cycle list shift=1,
            ]
            \addplot table [x=max_fork, y=Threshold, only if all={tie_break_mode/random}] {data/max_fork_bitcoin_whales.csv};

            \addplot table [x=max_fork, y=Threshold, only if all={tie_break_mode/random; ledger_function/longest}] {data/max_fork_whales.csv};

            \addplot table [x=max_fork, y=Threshold, only if all={tie_break_mode/random; ledger_function/mad}] {data/max_fork_whales.csv};
            \end{axis}
        \end{tikzpicture}%
        \label{fig:max-fork-lf:random}}
    \hfil
    \subfloat[Worst-case tie-breaking.]{%
        \begin{tikzpicture}
            \pgfplotsset{height=0.18\textwidth,width=0.33\textwidth}
            \begin{axis}[
                xlabel={Max Fork~$L$},
                ylabel={Security Threshold},
                addplot/.style={draw opacity=0.8},
                ymin=-0.05,
                ymax=0.45,
                cycle list shift=1,
            ]
            \addplot table [x=max_fork, y=Threshold, only if all={tie_break_mode/attacker}] {data/max_fork_bitcoin_whales.csv};

            \addplot table [x=max_fork, y=Threshold, only if all={tie_break_mode/attacker; ledger_function/longest}] {data/max_fork_whales.csv};

            \addplot table [x=max_fork, y=Threshold, only if all={tie_break_mode/attacker; ledger_function/mad}] {data/max_fork_whales.csv};
            \end{axis}
        \end{tikzpicture}%
        \label{fig:max-fork-lf:attacker}}
    \\
    \vspace*{0.5em}
    \pgfplotslegendfromname{max-fork-lf-legend}

    \caption{Security threshold of NC, Canonical-DAG, and MAD-DAG as a function of maximum fork length~$L$.}
    \label{fig:max-fork-lf}
    \Description{Security threshold of NC, Canonical-DAG, and MAD-DAG as a function of maximum fork length~$L$.}
\end{figure*}

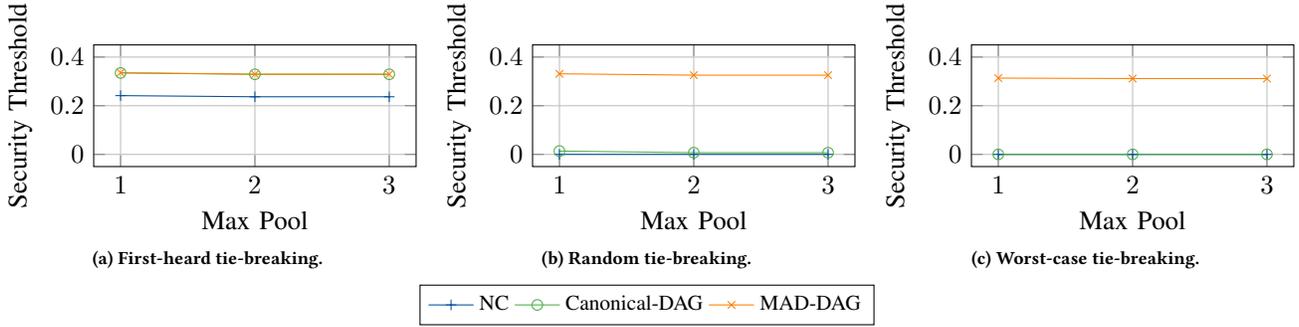
\begin{figure*}[t]
    \centering
    \subfloat[First-heard tie-breaking.]{%
        \begin{tikzpicture}
            \pgfplotsset{height=0.18\textwidth,width=0.33\textwidth}
            \begin{axis}[
                xlabel={Max Pool},
                ylabel={Security Threshold},
                legend pos=south east,
                addplot/.style={draw opacity=0.8},
                ymin=-0.05,
                ymax=0.45,
                cycle list shift=1,
                legend columns=3,
                legend to name=max-pool-legend
            ]
            \addplot table [x=max_pool, y=Threshold, only if all={tie_break_mode/first_heard; fee/2}] {data/bitcoin_max_pool.csv};
            \addlegendentry{NC}

            \addplot table [x=max_pool, y=Threshold, only if all={tie_break_mode/first_heard; ledger_function/longest}] {data/max_pool.csv};
            \addlegendentry{Canonical-DAG}

            \addplot table [x=max_pool, y=Threshold, only if all={tie_break_mode/first_heard; ledger_function/mad}] {data/max_pool.csv};
            \addlegendentry{MAD-DAG}
            \end{axis}
        \end{tikzpicture}%
        \label{fig:max-pool:first-heard}}
    \hfil
    \subfloat[Random tie-breaking.]{%
        \begin{tikzpicture}
            \pgfplotsset{height=0.18\textwidth,width=0.33\textwidth}
            \begin{axis}[
                xlabel={Max Pool},
                ylabel={Security Threshold},
                addplot/.style={draw opacity=0.8},
                ymin=-0.05,
                ymax=0.45,
                cycle list shift=1,
            ]
            \addplot table [x=max_pool, y=Threshold, only if all={tie_break_mode/random; fee/2}] {data/bitcoin_max_pool.csv};

            \addplot table [x=max_pool, y=Threshold, only if all={tie_break_mode/random; ledger_function/longest}] {data/max_pool.csv};

            \addplot table [x=max_pool, y=Threshold, only if all={tie_break_mode/random; ledger_function/mad}] {data/max_pool.csv};
            \end{axis}
        \end{tikzpicture}%
        \label{fig:max-pool:random}}
    \hfil
    \subfloat[Worst-case tie-breaking.]{%
        \begin{tikzpicture}
            \pgfplotsset{height=0.18\textwidth,width=0.33\textwidth}
            \begin{axis}[
                xlabel={Max Pool},
                ylabel={Security Threshold},
                addplot/.style={draw opacity=0.8},
                ymin=-0.05,
                ymax=0.45,
                cycle list shift=1,
            ]
            \addplot table [x=max_pool, y=Threshold, only if all={tie_break_mode/attacker; fee/2}] {data/bitcoin_max_pool.csv};

            \addplot table [x=max_pool, y=Threshold, only if all={tie_break_mode/attacker; ledger_function/longest}] {data/max_pool.csv};

            \addplot table [x=max_pool, y=Threshold, only if all={tie_break_mode/attacker; ledger_function/mad}] {data/max_pool.csv};
            \end{axis}
        \end{tikzpicture}%
        \label{fig:max-pool:attacker}}
    \\
    \vspace*{0.5em}
    \pgfplotslegendfromname{max-pool-legend}

    \caption{Security threshold of NC, Canonical-DAG, and MAD-DAG as a function of maximum pool size.}
    \label{fig:max-pool}
    \Description{Security threshold of NC, Canonical-DAG, and MAD-DAG as a function of maximum pool size.}
\end{figure*}

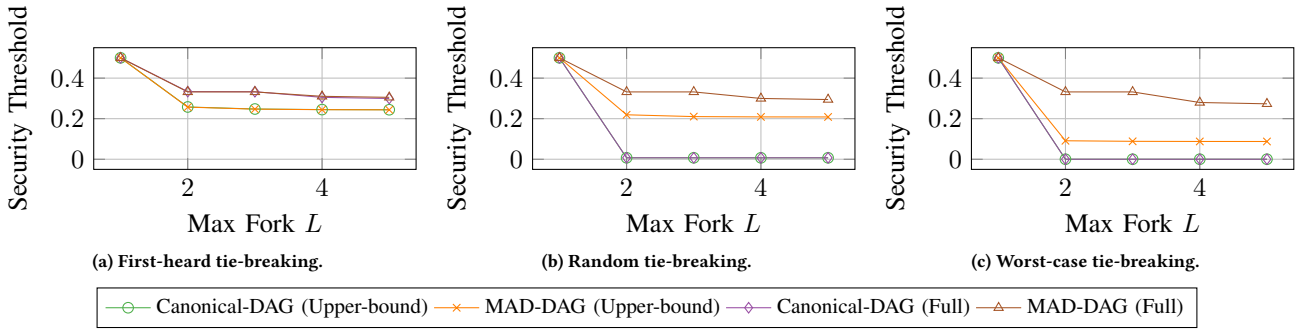
\begin{figure*}[t]
    \centering
    \subfloat[First-heard tie-breaking.]{%
        \begin{tikzpicture}
            \pgfplotsset{height=0.18\textwidth,width=0.33\textwidth}
            \begin{axis}[
                xlabel={Max Fork~$L$},
                ylabel={Security Threshold},
                legend pos=south east,
                addplot/.style={draw opacity=0.8},
                ymin=-0.05,
                ymax=0.55,
                cycle list shift=2,
                legend columns=4,
                legend to name=max-fork-lf-comparison-legend
            ]
            \addplot table [x=max_fork, y=Threshold, only if all={tie_break_mode/first_heard; difficulty_source/main; ledger_function/longest; model/simplified_colordag}] {data/colordag_model_comparison_max_fork_with_fees.csv};
            \addlegendentry{Canonical-DAG (Upper-bound)}

            \addplot table [x=max_fork, y=Threshold, only if all={tie_break_mode/first_heard; difficulty_source/main; ledger_function/mad; model/simplified_colordag}] {data/colordag_model_comparison_max_fork_with_fees.csv};
            \addlegendentry{MAD-DAG (Upper-bound)}
            
            \addplot table [x=max_fork, y=Threshold, only if all={tie_break_mode/first_heard; difficulty_source/main; ledger_function/longest; model/chain_colordag}] {data/colordag_model_comparison_max_fork_with_fees.csv};
            \addlegendentry{Canonical-DAG (Full)}

            \addplot table [x=max_fork, y=Threshold, only if all={tie_break_mode/first_heard; difficulty_source/main; ledger_function/mad; model/chain_colordag}] {data/colordag_model_comparison_max_fork_with_fees.csv};
            \addlegendentry{MAD-DAG (Full)}
            \end{axis}
        \end{tikzpicture}%
        \label{fig:max-fork-lf-comparison:first-heard}}
    \hfil
    \subfloat[Random tie-breaking.]{%
        \begin{tikzpicture}
            \pgfplotsset{height=0.18\textwidth,width=0.33\textwidth}
            \begin{axis}[
                xlabel={Max Fork~$L$},
                ylabel={Security Threshold},
                addplot/.style={draw opacity=0.8},
                ymin=-0.05,
                ymax=0.55,
                cycle list shift=2,
            ]
            \addplot table [x=max_fork, y=Threshold, only if all={tie_break_mode/random; difficulty_source/main; ledger_function/longest; model/simplified_colordag}] {data/colordag_model_comparison_max_fork_with_fees.csv};

            \addplot table [x=max_fork, y=Threshold, only if all={tie_break_mode/random; difficulty_source/main; ledger_function/mad; model/simplified_colordag}] {data/colordag_model_comparison_max_fork_with_fees.csv};

            \addplot table [x=max_fork, y=Threshold, only if all={tie_break_mode/random; difficulty_source/main; ledger_function/longest; model/chain_colordag}] {data/colordag_model_comparison_max_fork_with_fees.csv};

            \addplot table [x=max_fork, y=Threshold, only if all={tie_break_mode/random; difficulty_source/main; ledger_function/mad; model/chain_colordag}] {data/colordag_model_comparison_max_fork_with_fees.csv};
            \end{axis}
        \end{tikzpicture}%
        \label{fig:max-fork-lf-comparison:random}}
    \hfil
    \subfloat[Worst-case tie-breaking.]{%
        \begin{tikzpicture}
            \pgfplotsset{height=0.18\textwidth,width=0.33\textwidth}
            \begin{axis}[
                xlabel={Max Fork~$L$},
                ylabel={Security Threshold},
                addplot/.style={draw opacity=0.8},
                ymin=-0.05,
                ymax=0.55,
                cycle list shift=2,
            ]
            \addplot table [x=max_fork, y=Threshold, only if all={tie_break_mode/attacker; difficulty_source/main; ledger_function/longest; model/simplified_colordag}] {data/colordag_model_comparison_max_fork_with_fees.csv};

            \addplot table [x=max_fork, y=Threshold, only if all={tie_break_mode/attacker; difficulty_source/main; ledger_function/mad; model/simplified_colordag}] {data/colordag_model_comparison_max_fork_with_fees.csv};

            \addplot table [x=max_fork, y=Threshold, only if all={tie_break_mode/attacker; difficulty_source/main; ledger_function/longest; model/chain_colordag}] {data/colordag_model_comparison_max_fork_with_fees.csv};

            \addplot table [x=max_fork, y=Threshold, only if all={tie_break_mode/attacker; difficulty_source/main; ledger_function/mad; model/chain_colordag}] {data/colordag_model_comparison_max_fork_with_fees.csv};
            \end{axis}
        \end{tikzpicture}%
        \label{fig:max-fork-lf-comparison:attacker}}
    \\
    \vspace*{0.5em}
    \pgfplotslegendfromname{max-fork-lf-comparison-legend}

    \caption{Comparison of security threshold,  Canonical-DAG, and MAD-DAG in the simplified model and the full model as a function of maximum fork length~$L$.}
    \label{fig:max-fork-lf-comparison}
    \Description{Comparison of security threshold,  Canonical-DAG, and MAD-DAG in the simplified model and the full model as a function of maximum fork length~$L$.}
\end{figure*}

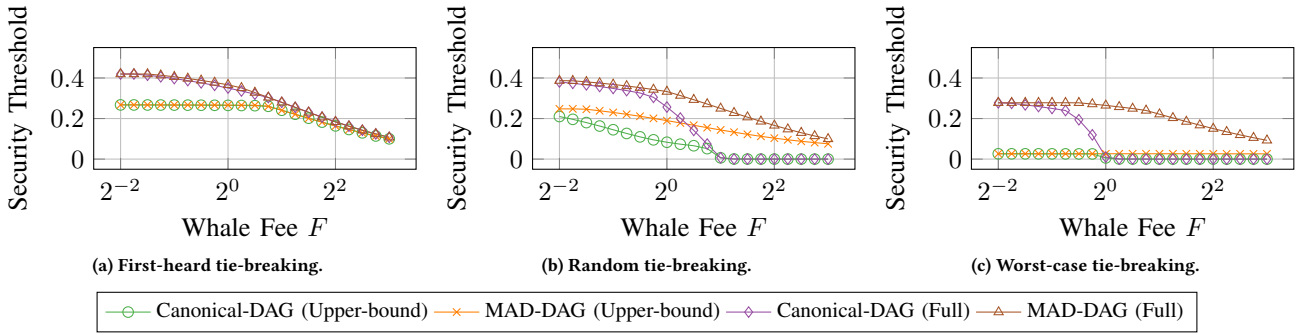
\begin{figure*}[t]
    \centering
    \subfloat[First-heard tie-breaking.]{%
        \begin{tikzpicture}
            \pgfplotsset{height=0.18\textwidth,width=0.33\textwidth}
            \begin{axis}[
                xlabel={Whale Fee~$F$},
                ylabel={Security Threshold},
                legend pos=south east,
                addplot/.style={draw opacity=0.8},
                ymin=-0.05,
                ymax=0.55,
                xmode=log,
                log basis x=2,
                cycle list shift=2,
                legend columns=4,
                legend to name=whale-fee-lf-comparison-legend
            ]
            \addplot table [x=fee, y=Threshold, only if all={tie_break_mode/first_heard; difficulty_source/main; ledger_function/longest; model/simplified_colordag}] {data/colordag_model_comparison_F_with_fees.csv};
            \addlegendentry{Canonical-DAG (Upper-bound)}

            \addplot table [x=fee, y=Threshold, only if all={tie_break_mode/first_heard; difficulty_source/main; ledger_function/mad; model/simplified_colordag}] {data/colordag_model_comparison_F_with_fees.csv};
            \addlegendentry{MAD-DAG (Upper-bound)}
            
            \addplot table [x=fee, y=Threshold, only if all={tie_break_mode/first_heard; difficulty_source/main; ledger_function/longest; model/chain_colordag}] {data/colordag_model_comparison_F_with_fees.csv};
            \addlegendentry{Canonical-DAG (Full)}

            \addplot table [x=fee, y=Threshold, only if all={tie_break_mode/first_heard; difficulty_source/main; ledger_function/mad; model/chain_colordag}] {data/colordag_model_comparison_F_with_fees.csv};
            \addlegendentry{MAD-DAG (Full)}
            \end{axis}
        \end{tikzpicture}%
        \label{fig:whale-fee-lf-comparison:first-heard}}
    \hfil
    \subfloat[Random tie-breaking.]{%
        \begin{tikzpicture}
            \pgfplotsset{height=0.18\textwidth,width=0.33\textwidth}
            \begin{axis}[
                xlabel={Whale Fee~$F$},
                ylabel={Security Threshold},
                addplot/.style={draw opacity=0.8},
                ymin=-0.05,
                ymax=0.55,
                xmode=log,
                log basis x=2,
                cycle list shift=2,
            ]
            \addplot table [x=fee, y=Threshold, only if all={tie_break_mode/random; difficulty_source/main; ledger_function/longest; model/simplified_colordag}] {data/colordag_model_comparison_F_with_fees.csv};

            \addplot table [x=fee, y=Threshold, only if all={tie_break_mode/random; difficulty_source/main; ledger_function/mad; model/simplified_colordag}] {data/colordag_model_comparison_F_with_fees.csv};

            \addplot table [x=fee, y=Threshold, only if all={tie_break_mode/random; difficulty_source/main; ledger_function/longest; model/chain_colordag}] {data/colordag_model_comparison_F_with_fees.csv};

            \addplot table [x=fee, y=Threshold, only if all={tie_break_mode/random; difficulty_source/main; ledger_function/mad; model/chain_colordag}] {data/colordag_model_comparison_F_with_fees.csv};
            \end{axis}
        \end{tikzpicture}%
        \label{fig:whale-fee-lf-comparison:random}}
    \hfil
    \subfloat[Worst-case tie-breaking.]{%
        \begin{tikzpicture}
            \pgfplotsset{height=0.18\textwidth,width=0.33\textwidth}
            \begin{axis}[
                xlabel={Whale Fee~$F$},
                ylabel={Security Threshold},
                addplot/.style={draw opacity=0.8},
                ymin=-0.05,
                ymax=0.55,
                xmode=log,
                log basis x=2,
                cycle list shift=2,
            ]
            \addplot table [x=fee, y=Threshold, only if all={tie_break_mode/attacker; difficulty_source/main; ledger_function/longest; model/simplified_colordag}] {data/colordag_model_comparison_F_with_fees.csv};

            \addplot table [x=fee, y=Threshold, only if all={tie_break_mode/attacker; difficulty_source/main; ledger_function/mad; model/simplified_colordag}] {data/colordag_model_comparison_F_with_fees.csv};

            \addplot table [x=fee, y=Threshold, only if all={tie_break_mode/attacker; difficulty_source/main; ledger_function/longest; model/chain_colordag}] {data/colordag_model_comparison_F_with_fees.csv};

            \addplot table [x=fee, y=Threshold, only if all={tie_break_mode/attacker; difficulty_source/main; ledger_function/mad; model/chain_colordag}] {data/colordag_model_comparison_F_with_fees.csv};
            \end{axis}
        \end{tikzpicture}%
        \label{fig:whale-fee-lf-comparison:attacker}}
    \\
    \vspace*{0.5em}
    \pgfplotslegendfromname{whale-fee-lf-comparison-legend}

    \caption{Comparison of security threshold, Canonical-DAG, and MAD-DAG in the simplified model and the full model as a function of whale fee~$F$.}
    \label{fig:whale-fee-lf-comparison}
    \Description{Comparison of security threshold, Canonical-DAG, and MAD-DAG in the simplified model and the full model as a function of whale fee~$F$.}
\end{figure*}

%%%%%%%%%%%%%%%%%%%%%%%%%%%%%%%%%%%%%%%%%%%%%%%%%%%%%%%%%%%%%%%%%%%%%%%%%%%%%%%%%%%%%%%%%%%%%%%%%%%%%%%%%%%%%%%%%%%%%%%%%%%%%%%%%%%%%%%%%%%%%%%%%%%%%%%%

We begin by analyzing the effect of the difficulty-adjustment mechanism.
We compare (1) NC, (2) Colordag, which uses the uncontested difficulty-adjustment mechanism, and (3) \emph{Canonical-DAG}, a protocol similar to Colordag that uses the canonical difficulty-adjustment mechanism instead.
All of these protocols use the canonical ledger function.

Canonical-DAG serves as a stepping stone to indirectly compare Colordag and MAD-DAG while isolating the effects of the different mechanisms on the protocol.
First, in this section, we compare Colordag and Canonical-DAG to show the latter is more secure due to the use of the canonical difficulty-adjustment mechanism.
Then, in the next section, we will compare Canonical-DAG and MAD-DAG to show the latter is more secure due to the use of the MAD ledger function, and consequently, is also more secure compared to Colordag.

We analyze Colordag and Canonical-DAG using the upper-bound model~(\quickSectionRef{sec:analyzing-selfish-mining:upper-bound-model}).
We analyze NC using a model similar to that of Bar-Zur et al.~\cite{bar2022werlman}~(Appendix~\ref{appendix:nc-selfish-mining-model}), but with the same method for updating the \pool that we use in the new models.

While our method only allows us to compare bounds on the security threshold, it is still extremely valuable:
These are the first bounds provided under adverse conditions, and a higher lower-bound on the security threshold is preferable.

Since we focus on the difficulty-adjustment mechanism in this section, we ignore whale transactions and set~${\delta = 0}$.

All comparisons are done for all three tie-breaking rules.
For the first-heard tie-breaking rule, we use a rushing factor~$\gamma$ of~0.5, as often done in previous work~\cite{sapirshtein2017optimal,hou2019squirrl,bar2022werlman}.

\paragraph{Revenue}

We first analyze the revenue of a selfish miner in NC, Colordag, and Canonical-DAG as a function of the selfish miner's size~$\alpha$ (Figure~\ref{fig:revenue-dam}).
We set~$\forkSensitivity$ to 15 and \maxFork to 10 blocks.
We plot the revenue for each of the three tie-breaking rules.

We observe that a selfish miner in Canonical-DAG obtains less revenue compared to both NC and Colordag.
This experiment also demonstrates that the selfish miner's revenue is equal to their \emph{fair share} (how much could be obtained by mining honestly) until a critical point (the security threshold), after which selfish mining yields more revenue than honest mining.

\paragraph{Fork Sensitivity}

We focus next on calculating the security threshold rather than only revenue.
We compare the security threshold of the three protocols as a function of~$\forkSensitivity$ (Figure~\ref{fig:threshold-dam}).
In this experiment, we set~$\maxFork$ to 20.

The security threshold of both Colordag and Canonical-DAG increases with~$\forkSensitivity$.
This is in line with the analysis of Colordag~\cite{abraham2023colordag}.
While low values of~$\forkSensitivity$ result with Colordag having a lower security threshold than NC for the first-heard and random tie-breaking rules, this is quickly reversed for higher values of~$\forkSensitivity$, as Colordag's subsidy mechanism becomes more effective as~$\forkSensitivity$ increases.

Canonical-DAG has a higher security threshold compared to NC and Colordag for all tie-breaking rules and all values of~$\forkSensitivity$.

\paragraph{Guaranteed fee}

Next, we analyze how the guaranteed fee~$f$ (a baseline level of fees that all blocks in the canonical chain receive) affects the security threshold (Figure~\ref{fig:guaranteed-fee-dam}).
We set~$\forkSensitivity$ to 25 and~$\maxFork$ to 10.

Again, Canonical-DAG has a higher security threshold compared to NC and Colordag in all cases.

Colordag, however, has a lower security threshold compared to NC when~$f$ is high.
Increasing~$f$ shifts the revenue of miners from subsidy to the fees they obtain by creating blocks in the canonical chain regardless of whether they are contested or not.
This results in a reward structure similar to NC.
However, in Colordag, a selfish miner can still benefit from uncontested blocks due to them not contributing to the difficulty-adjustment mechanism, lowering the block creation difficulty and consequently increasing the rate of block creation.

In NC, the selfish miner objective is simply the original one (the fraction of their blocks in the canonical chain) multiplied by a constant.
Thus, this the guaranteed fee does not affect on the security threshold.

While the security threshold of Canonical-DAG decreases as~$f$ increases, similarly to Colordag, its security threshold approaches NC's as~$f$ increases due to the fact that it has the same difficulty-adjustment mechanism as NC.

\paragraph{State Space Restriction}

Next, we analyze how the maximum fork length~$\maxFork$ affects the security threshold (Figure~\ref{fig:max-fork-dam}).
We set~$\forkSensitivity$ to 25.

For all protocols, increasing \maxFork reduces the security threshold.
This effect becomes negligible when \maxFork is 10 or more.

\paragraph{Comparison to Full model}

Last, we compare the results between the full model and the upper-bound model for both Colordag and Canonical-DAG~(Figure~\ref{fig:max-fork-dam-comparison}).
We vary \maxFork from~1 to~5, as the full model is more complex and requires too much memory for higher values.
We also set \forkSensitivity to 5.

As expected, across all protocols and tie-breaking rules, increasing \maxFork reduces the security threshold, as it allows more elaborate strategies for the selfish miner.
In addition, the results corroborate the correctness of the upper-bound model:
It has a lower security threshold compared to the full model for both protocols, which is expected since it uses an upper bound on the revenue of the selfish miner.

The full model analysis of Canonical-DAG also yields a lower security threshold compared to Colordag, similarly to the upper-bound model.

%%%%%%%%%%%%%%%%%%%%%%%%%%%%%%%%%%%%%%%%%%%%%%%%%%%%%%%%%%%%%%%%%%%%%%%%%%%%%%%%%%%%%%%%%%%%%%%%%%%%%%%%%%%%%%%%%%%%%%%%%%%%%%%%%%%%%%%%%%%%%%%%%%%%%%%%%%%%%%%%%%%%%%%%%%%%%%%%%%%%%%%%%%%%%%%%%%%%%%%%%%%%%%%%%%%%%%%%%%%%%%%%%%%%%%%%%%%%%%%%%%%%%%%%%%%%%%%%%%%%%%%%%%%%%%%%%%%%%%%%%%%%%%%%%%%%%%%%%%%%%%%%%%%%%%%%%%%%%%%%%%%%%%%%%%%%%%%%%%%%%%%%%%%%%%%%%%%%%%%%%%%%%%%%%%%%%%%%%%%%%%%%%%%%%%%%%%%%%%%%%%%%%%%%%%%%%%%%%%%%%%%%%%%%%%%%%%%%%%%%%%%%%%%%%%%%%%%%

\section{Ledger Function}\label{sec:ledger-function}

%%%%%%%%%%%%%%%%%%%%%%%%%%%%%%%%%%%%%%%%%%%%%%%%%%%%%%%%%%%%%%%%%%%%%%%%%%%%%%%%%%%%%%%%%%%%%%%%%%%%%%%%%%%%%%%%%%%%%%%%%%%%%%%%%%%%%%%%%%%%%%%%%%%%%%%%%%%%%%%%%%%%%%%%%%%%%%%%%%%%%%%%%%%%%%%%%%%%%%%%%%%%%%%%%%%%%%%%%%%%%%%%%%%%%%%%%%%%%%%%%%%%%%%%%%%%%%%%%%%%%%%%%%%%%%%%%%%%%%%%%%%%%%%%%%%%%%%%%%%%%%%%%%%%%%%%%%%%%%%%%%%%%%%%%%%%%%%%%%%%%%%%%%%%%%%%%%%%%%%%%%%%%%%%%%%%%%%%%%%%%%%%%%%%%%%%%%%%%%%%%%%%%%%%%%%%%%%%%%%%%%%%%%%%%%%%%%%%%%%%%%%%%%%%%%%%%%%%

Having established that the Canonical-DAG, which uses the canonical difficulty-adjustment mechanism, is more secure than Colordag, which uses the uncontested difficulty-adjustment mechanism, we move on to analyze the impact of the ledger function.
To isolate the impact of the ledger function, we compare in this section the security of NC, Canonical-DAG, and MAD-DAG.
Showing that MAD-DAG is more secure that Canonical-DAG would lead us to conclude that it is more secure than Colordag as well.

For understanding the impact of the ledger function, in this section, we consider whale transactions and set the whale frequency~$\delta$ to~$0.01$, the whale transaction fee~$F$ to~$2$, and the maximum number of concurrent pending whale transactions~\maxPool to~2 unless stated otherwise.

We analyze Canonical-DAG and MAD-DAG with the upper-bound model~(\quickSectionRef{sec:analyzing-selfish-mining:upper-bound-model}), this time with the canonical difficulty-adjustment mechanism mode.
And again, we analyze NC using a model similar to the one by Bar-Zur et al.~\cite{bar2022werlman}~(Appendix~\ref{appendix:nc-selfish-mining-model}).

Again, we use~${\gamma = 0.5}$ for the first-heard tie-breaking rule.

We also fix~${\forkSensitivity = 15}$ and~${\maxFork = 10}$ unless otherwise stated.

\paragraph{Revenue}

We first analyze the revenue of a selfish miner in NC, Canonical-DAG, and MAD-DAG as a function of the selfish miner's size~$\alpha$ (Figure~\ref{fig:revenue-lf}).

For all miner sizes~$\alpha$ and tie-breaking rules we consider, a selfish miner can get more revenue in NC compared to Canonical-DAG and more revenue in Canonical-DAG compared to MAD-DAG.

We again, move on to compare the security of the three protocols.

\paragraph{Fork sensitivity}

We study the security threshold as a function of~$\forkSensitivity$ (Figure~\ref{fig:fork-sensitivity-lf}).
In line with the previous experiment, in all cases we consider, Canonical-DAG has a higher security threshold compared to NC and MAD-DAG has a higher security threshold compared to Canonical-DAG.

With the random and worst-case tie-breaking rules, the security threshold of Canonical-DAG is near-zero for all values of~$\forkSensitivity$, while the security threshold of MAD-DAG is significantly higher, above 30\% for~$\forkSensitivity \geq 15$.

\paragraph{Whale transactions}

Next, we compare the security threshold of the three protocols as a function of the whale transaction fee~$F$ (Figure~\ref{fig:whale-fee-lf}).
We again observe that the security threshold is highest for MAD-DAG, followed by Canonical-DAG, and lowest for NC.

In all protocols, the security threshold decreases when~$F$ increases.
This is in line with previous work that shows high whale fees increase selfish mining profitability~\cite{bar2022werlman,bar2023deep}.

With the first-heard tie-breaking rule (with~${\gamma = 0.5}$), the security thresholds of Canonical-DAG and MAD-DAG are similar, meaning the MAD ledger function has little  effect in this case.
With the random and worst-case tie-breaking rules, the security threshold of Canonical-DAG drops significantly starting from~${F=1}$ and~${F=0.7}$, to near-0 at~${F=2}$ and~${F=1}$, respectively.
This is because in these cases, when an honest miner creates a block with a whale transaction, it is more profitable for the selfish miner to fight for the whale transaction by attempting to create and to publish an alternative chain with a single block holding that whale transaction.
MAD-DAG's security threshold is largely unaffected by the choice of the tie-breaking rule and while it decreases for high values of~$F$, it remains significantly higher than the alternatives: 19.9\% when~${F=4}$ and 11\% when~${F=8}$.
This is because the previously described strategy does not work when the MAD ledger function is employed.

\paragraph{State Space Restriction}

% Max fork
We next analyze how the maximum fork length~$\maxFork$ affects the security threshold (Figure~\ref{fig:max-fork-lf}).
As before, increasing~\maxFork reduces the security threshold of all protocols, but the threshold remains virtually the same when~$\maxFork \geq 10$.

% Max pool
We also analyze how the maximum pool size~$\maxPool$ affects the security threshold (Figure~\ref{fig:max-pool}).
In all protocols, increasing~\maxPool from~1 to~2 slightly reduces the security threshold, and increasing it further from~2 to~3 has a negligible effect.
This is since whale transactions are rare~($\delta = 0.01$) and are unlikely to be simultaneously available.

\paragraph{Comparison to Full model}

Last, we compare the differences in security threshold between the upper-bound model and the full model for Canonical-DAG and MAD-DAG.

We first analyze the effect of the maximum fork length~$\maxFork$ on the security threshold (Figure~\ref{fig:max-fork-lf-comparison}).
Since the full model is more complex, we set~${\forkSensitivity = 5}$ and vary~$\maxFork$ from~1 to~5.

As in the previous comparison to the full model, the fact that for both protocols the security threshold in the upper-bound model is lower compared to the full model corroborates the correctness of the upper-bound model.
The security threshold decreases again when~$\maxFork$ increases, as expected.

While with the first-heard tie-breaking rule, there is little difference between the two protocols, with the random and worst-case tie-breaking rules, the security threshold of MAD-DAG is significantly higher than Canonical-DAG in both models.

The large difference between the threshold of MAD-DAG in the full model and the upper-bound model suggests that MAD-DAG may be significantly more secure than our analysis can guarantee.

Next, we analyze the effect of the whale fee~$F$ on the security threshold (Figure~\ref{fig:whale-fee-lf-comparison}).
This time, we set~${\maxFork = 3}$ and~${\forkSensitivity = 3}$.

Similar findings to the previous experiment emerge here:
The security threshold decreases when~$F$ increases;
with the first-heard tie-breaking rule the two protocols have similar security thresholds, while with the other rules, MAD-DAG is significantly more secure;
and there is a large difference between the security threshold of MAD-DAG that the upper-bound model predicts to the one the full model predicts, especially when~$F$ is low, meaning that the upper-bound model is not tight and can perhaps be improved in future work.

%%%%%%%%%%%%%%%%%%%%%%%%%%%%%%%%%%%%%%%%%%%%%%%%%%%%%%%%%%%%%%%%%%%%%%%%%%%%%%%%%%%%%%%%%%%%%%%%%%%%%%%%%%%%%%%%%%%%%%%%%%%%%%%%%%%%%%%%%%%%%%%%%%%%%%%%%%%%%%%%%%%%%%%%%%%%%%%%%%%%%%%%%%%%%%%%%%%%%%%%%%%%%%%%%%%%%%%%%%%%%%%%%%%%%%%%%%%%%%%%%%%%%%%%%%%%%%%%%%%%%%%%%%%%%%%%%%%%%%%%%%%%%%%%%%%%%%%%%%%%%%%%%%%%%%%%%%%%%%%%%%%%%%%%%%%%%%%%%%%%%%%%%%%%%%%%%%%%%%%%%%%%%%%%%%%%%%%%%%%%%%%%%%%%%%%%%%%%%%%%%%%%%%%%%%%%%%%%%%%%%%%%%%%%%%%%%%%%%%%%%%%%%%%%%%%%%%%%

\section{Conclusion}\label{sec:conclusion}

%%%%%%%%%%%%%%%%%%%%%%%%%%%%%%%%%%%%%%%%%%%%%%%%%%%%%%%%%%%%%%%%%%%%%%%%%%%%%%%%%%%%%%%%%%%%%%%%%%%%%%%%%%%%%%%%%%%%%%%%%%%%%%%%%%%%%%%%%%%%%%%%%%%%%%%%%%%%%%%%%%%%%%%%%%%%%%%%%%%%%%%%%%%%%%%%%%%%%%%%%%%%%%%%%%%%%%%%%%%%%%%%%%%%%%%%%%%%%%%%%%%%%%%%%%%%%%%%%%%%%%%%%%%%%%%%%%%%%%%%%%%%%%%%%%%%%%%%%%%%%%%%%%%%%%%%%%%%%%%%%%%%%%%%%%%%%%%%%%%%%%%%%%%%%%%%%%%%%%%%%%%%%%%%%%%%%%%%%%%%%%%%%%%%%%%%%%%%%%%%%%%%%%%%%%%%%%%%%%%%%%%%%%%%%%%%%%%%%%%%%%%%%%%%%%%%%%%%

We present MAD-DAG, the first practical DAG-based blockchain protocol that disincentivizes selfish mining under adverse conditions including rushing attacks, varying block rewards, and petty-compliant miners.
This is achieved through the novel MAD ledger function that discards content from equal-length competing chains.
We conduct the first tractable MDP-based analysis of selfish mining in DAG-based blockchains using a novel upper-bound model that conservatively favors the selfish miner.

Our analysis demonstrates that MAD-DAG's security threshold consistently matches or exceeds Colordag's and NC's.
Critically, MAD-DAG retains its security against well-connected selfish miners and petty-compliant miners, while selfish mining becomes profitable in both NC and Colordag for miners of any size.

Furthermore, MAD-DAG achieves these improvements with fork sensitivity as low as~15, enabling practical deployment unlike Colordag which requires impractically high values.

As Bitcoin's block subsidies halve every four years and projects like BitVM introduce reward variability, Bitcoin faces escalating vulnerability to selfish mining.
MAD-DAG provides the most robust defense available.

\begin{acks}
This work was supported by research grants from the Avalanche foundation, and IC3, the Initiative for CryptoCurrencies and Contracts.
This work also received funding from the European Union (ERC, Bayes-RL, Project Number 101041250).
Views and opinions expressed are however those of the authors only and do not necessarily reflect those of the European Union or the European Research Council Executive Agency.
Neither the European Union nor the granting authority can be held responsible for them.
\end{acks}

\bibliographystyle{ACM-Reference-Format}
\balance
\bibliography{references}

\appendix

%%%%%%%%%%%%%%%%%%%%%%%%%%%%%%%%%%%%%%%%%%%%%%%%%%%%%%%%%%%%%%%%%%%%%%%%%%%%%%%%%%%%%%%%%%%%%%%%%%%%%%%%%%%%%%%%%%%%%%%%%%%%%%%%%%%%%%%%%%%%%%%%%%%%%%%%

\section{Calculating the Average Number of Whale Transactions per Block}
\label{appendix:average-whales-included}

Assuming all miners are honest, it means that whenever a new block is created, it includes a whale transaction if one is available and that the new block is immediately published and accepted by everyone else.
Also, recall that when the number of whale transactions is~\maxPool, any excess transactions are discarded.

The number of whale transactions follows a Markov chain with states~$0, 1, ..., \maxPool$.
For~$1 < i < \maxPool$, the probability to move right (the number of transactions increases by~1) is~$\frac{\delta}{1 + \delta}$ and the probability to move left (the number of transactions decreases by~1) is~$\frac{1}{1 + \delta}$ (this matches the ratio~$\delta$ that our model stipulates).
For~$i = 0$, the probability to move right is~$\frac{\delta}{1 + \delta}$ and the probability to stay is~$\frac{1}{1 + \delta}$.
For~$i = \maxPool$, the probability to stay is~$\frac{\delta}{1 + \delta}$ and the probability to move left it~$\frac{1}{1 + \delta}$.

Calculating the steady state distribution of the chain is straightforward.
We find that the probability to be in state~0 (no whale transactions are available) is exactly~${p_0 = \frac{1 - \delta}{1 - \delta^{\maxPool + 1}}}$.
The number of average whale transactions per block is its complement as a whale transaction will be included whenever one is available, resulting in~${q = \frac{\delta - \delta^{\maxPool + 1}}{1 - \delta^{\maxPool + 1}}}$.

%%%%%%%%%%%%%%%%%%%%%%%%%%%%%%%%%%%%%%%%%%%%%%%%%%%%%%%%%%%%%%%%%%%%%%%%%%%%%%%%%%%%%%%%%%%%%%%%%%%%%%%%%%%%%%%%%%%%%%%%%%%%%%%%%%%%%%%%%%%%%%%%%%%%%%%%

\section{NC Selfish Mining Model}
\label{appendix:nc-selfish-mining-model}

We review the model of Bar-Zur et al.~\cite{bar2022werlman} of NC with varying rewards.
The model captures the perspective of a selfish miner who controls a fraction of ~$\alpha$ of the mining power while all other miners are honest.

The miner has a rushing factor~$\gamma$.

The subsidy is constant and equal to~$1$ unit.
Occasionally, a whale transaction appears, with a fee of~$F$ units.
Each block can contain a single transaction or none.

This model restricts the miner to mine on a single chain and does not track any blocks before the last fork.
This is because an optimal selfish miner strategy only requires the miner to mine on a single chain.
If at any point in time, a selfish miner with an optimal strategy switches to mine on another chain, it means it is more profitable to do so.
At that point, the chance that the old chain will be rewarded only decreases, leaving it a suboptimal choice.
This holds for all future steps as well, meaning the old chain should be abandoned, and the miner should focus only on the new chain.

%%%%%%%%%%%%%%%%%%%%%%%%%%%%%%%%%%%%%%%%%%%%%%%%%%%%%%%%%%%%%%%%%%%%%%%%%%%%%%%%%%%%%%%%%%%%%%%%%%%%%%%%%%%%%%%%%%%%%%%%%%%%%%%%%%%%%%%%%%%%%%%%%%%%%%%%

\subsubsection{Utility Function}

The utility function takes a similar form to the model of Sapirshtein et al.~\cite{sapirshtein2017optimal}: the expected ratio between the total reward and the total difficulty contribution.
The reward~$R_t$ includes subsidy as well as transaction fees that the selfish miner receives, while~$D_t$ remains the number of blocks added to the longest chain at step~$t$.

%%%%%%%%%%%%%%%%%%%%%%%%%%%%%%%%%%%%%%%%%%%%%%%%%%%%%%%%%%%%%%%%%%%%%%%%%%%%%%%%%%%%%%%%%%%%%%%%%%%%%%%%%%%%%%%%%%%%%%%%%%%%%%%%%%%%%%%%%%%%%%%%%%%%%%%%

\subsubsection{States}

The state space includes all states of the form~${\left( \aChain, \hChain, \fork, \pool \right)}$, where~$\aChain$ is a vector of all blocks in the selfish miner's secret chain, including a flag for each block indicating if the block contains a whale transaction;
$\hChain$ is a vector of all honest blocks in the public chain, the honest miners' chain starting from the fork that the selfish miner created (again, including a flag for each block indicating if the block contains a whale transaction);
$\fork$ is a flag taking one of three values: \forkIrrelevant, \forkRelevant, or \forkActive, denoting whether rushing is possible or currently occurring;
and $\pool$ is the number of whale transactions that have appeared since the last fork.

Figure~\ref{fig:state-space:nc} shows an example state.

%%%%%%%%%%%%%%%%%%%%%%%%%%%%%%%%%%%%%%%%%%%%%%%%%%%%%%%%%%%%%%%%%%%%%%%%%%%%%%%%%%%%%%%%%%%%%%%%%%%%%%%%%%%%%%%%%%%%%%%%%%%%%%%%%%%%%%%%%%%%%%%%%%%%%%%%

\subsubsection{Actions and Transitions}

The model allows the three following types of actions.
For convenience, we present these as parameterized actions that have a similar function, but in practice, each parameter value corresponds to a different action in the MDP.

\paragraph{Adopt~$\ell$}
The selfish miner abandons their secret chain, adopts the first~$\ell$ blocks of the public chain and prepares to mine on top of the~$\ell$-th block.
The pool is decremented by the number of whale transactions in the public chain~(${\pool \gets \pool - \transactions{\hChain}}$, where $\transactions{\cdot}$ is the number of whale transactions in the given chain).
In addition, a difficulty contribution of~$\ell$ is yielded.

\paragraph{Reveal~$\ell$}
This is only allowed when~$\ell \geq \abs{\aChain}$ (where~$\abs{\cdot}$ is the length of a vector) and either~$\ell > \abs{\hChain}$ (overriding the public chain) or~$\ell = \abs{\hChain}$ and~$\fork = \forkRelevant$ (attempting rushing).

In the first case, the selfish miner publishes exactly~$\ell$ blocks, causing all honest miners to discard the public chain and prepare to mine on top of the selfish miner's revealed chain.
The public chain is reset~(${\hChain \gets \emptyChain}$) and the selfish miner's secret chain shifts by~$\ell$ blocks~(${\aChain \gets \aChain \ll \ell}$).
The pool is decremented by the number of whale transactions in the revealed blocks~(${\pool \gets \pool - \transactions{\aChain[:\ell]}}$, where $\vec{x}[:n]$ is the prefix of $\vec{x}$ of length $n$).
A reward of~$\ell + F \cdot \transactions{\aChain[:\ell]}$ and a difficulty contribution of~$\ell$ are yielded.

In the second case, $\fork \gets \forkActive$.

\paragraph{Wait}
The selfish miner waits until a new block is created.

Denote by~$\vec{x}\increment$ the vector $\vec{x}$ with an appended block $x$ that contains a whale transaction if one can be added (depending on how many transactions~$x$ contains in comparison to~\pool) and 0 otherwise.

If \fork is not \forkActive, then one of two transitions may occur:
\begin{enumerate}
    \item The selfish miner mines a block in their secret chain~(${\aChain \gets \aChain\increment}$) with probability~$\alpha$, meaning rushing is currently impossible~(${\fork \gets \forkIrrelevant}$).
    \item An honest miner mines a block in the public chain~(${\hChain \gets \hChain\increment}$) with probability~$1-\alpha$, meaning rushing is currently possible~(${\fork \gets \forkRelevant}$).
\end{enumerate}

Otherwise, if \fork is \forkActive, there is rushing currently occurring.
Then, one of three transitions may occur:
\begin{enumerate}
    \item The selfish miner mines a block in their secret chain~(${\aChain \gets \aChain\increment}$) with probability~$\alpha$, meaning rushing is currently impossible~(${\fork \gets \forkIrrelevant}$);
    \item An honest miner who received the honest miner's block first mines a block in the public chain~(${\hChain \gets \hChain\increment}$) with probability~${(1 - \gamma) (1 - \alpha)}$.
    \item An honest miner who received the selfish miner's block first mines a block on the miner's chain with probability~${\gamma (1 - \alpha)}$; all miners now agree that the published part of the selfish miner's chain is in the longest chain, discarding the previous public chain in favor of the new honest block~(${\hChain \gets \emptyChain\increment}$), shortening the selfish miner's secret chain~(${\aChain \gets \aChain \ll \abs{\hChain}}$), and yielding a reward of~$\abs{\hChain} + F \cdot \transactions{\aChain[:\abs{\hChain}]}$ and a difficulty contribution of~$\abs{\hChain}$.
\end{enumerate}
In the two latter cases, further rushing is still possible (${\fork \gets \forkRelevant}$).

Whenever the longest chain's length increases, there is a chance of~$\delta$ that \pool is incremented by 1~(${\pool \gets \pool + 1}$).

\subsubsection{Bounding the State Space}

To numerically solve the MDP, the state space needs to be bounded.
The model allows the length of~$\aChain$ and~$\hChain$ to be up to a value of~$\maxFork$.
Whenever there is a possibility that a transition resulting from a certain action will violate this constraint, the action is forbidden.
It is important to confirm that we never reach a state with no possible actions, since then the MDP would not be valid.
But this is guaranteed, since a miner can always adopt the public chain to reduce the length of both chains.

In addition, the model also bounds the number of whale transactions in the pool by~$\maxPool$.
When more than~$\maxPool$ whale transactions are in the pool, the excess transactions are discarded.

\end{document}

%% file: style.tex
\usepackage{graphicx}
\usepackage{amsmath}
\usepackage[caption=false,font=footnotesize]{subfig}
\usepackage{url}

\usepackage{xstring}
\usepackage{amsfonts}

\usepackage{epstopdf}
\usepackage{soul}
\usepackage{mathtools}
\usepackage{tabularx,environ}
\usepackage{xspace}

\usepackage{hyperref}
\hypersetup{
  colorlinks   = true,    % Colours links instead of ugly boxes
  urlcolor     = blue,    % Colour for external hyperlinks
  linkcolor    = blue,    % Colour of internal links
  citecolor    = red      % Colour of citations
}

\usepackage{tikz}
\usetikzlibrary{external}
\usetikzlibrary{positioning,patterns,decorations.pathreplacing,calc,backgrounds}

\usepackage{pgfplots}
\usepackage{pgfplotstable}
\pgfplotsset{compat=1.18}
\usepgfplotslibrary[colorbrewer]
\usetikzlibrary{pgfplots.colorbrewer}

\pgfplotsset{
  only if/.style args={#1 is #2}{
    /pgfplots/x filter/.code={
      \edef\tempa{\thisrow{#1}}
      \edef\tempb{#2}
      \ifx\tempa\tempb
      \else
        
      \fi
    }
  },
  only if all/.style args={#1}{
    /pgfplots/x filter/.code={
      \edef\tempargs{#1}%
      \StrSubstitute{\tempargs}{;}{,}[\tempargs]%
      \foreach \col/\val in \tempargs {
        \edef\tempa{\thisrow{\col}}%
        \edef\tempb{\val}%
        \ifx\tempa\tempb
        \else
          \breakforeach
        \fi
      }
    }
  }
}

\newcommand{\pgfplotstablefindvalue}[5][]{%
    \def#5{#1}% set default
    \pgfplotstablegetcolumnnamebyindex{0}\of#2\to\@firstcol
    \pgfplotstableforeachcolumnelement{\@firstcol}\of#2\as\@dummy{%
        \def\@match{1}% assume match until proven otherwise
        \foreach \col/\val in {#3} {%
            \pgfplotstablegetelem{\pgfplotstablerow}{\col}\of#2
            \let\@cellvalue\pgfplotsretval
            \edef\@tempval{\val}%
            \ifx\@cellvalue\@tempval
            \else
                \global\def\@match{0}% no match
                \breakforeach
            \fi
        }%
        \ifnum\@match=1
            \pgfplotstablegetelem{\pgfplotstablerow}{#4}\of#2
            \global\let#5\pgfplotsretval
        \fi
    }%
}

\definecolor{color1}{HTML}{000000}
\definecolor{color2}{HTML}{08519c}
\definecolor{color3}{HTML}{4daf4a}
\definecolor{color4}{HTML}{ff7f00}
\definecolor{color5}{HTML}{984ea3}
\definecolor{color6}{HTML}{a65628}
\definecolor{color7}{HTML}{f781bf}
\definecolor{color8}{HTML}{ffff33}

\definecolor{colorNCrandom}{HTML}{08519c}
\definecolor{bluelight}{HTML}{3182bd}
\definecolor{bluelighter}{HTML}{6baed6}
\definecolor{bluelightest}{HTML}{bdd7e7}

\pgfplotsset{
  cycle multiindex* list={
    color1,color2,color3,color4,color5,color6,color7,color8
    \nextlist
    mark=square,mark=+,mark=o,mark=x,mark=diamond,mark=triangle
    \nextlist
  }
}

\pgfplotsset{
  /pgfplots/xmajorgrids=true,
  /pgfplots/ymajorgrids=true,
}

\pgfplotsset{every axis legend/.append style={font=\footnotesize}}

\pgfplotstableset{col sep=comma}

\newcommand{\quickSectionRef}[1]{\S\ref{#1}}

\newcommand{\forkSensitivity}{\ensuremath{N_\ell}\xspace}

\newcommand{\fork}{\textit{fork}\xspace}
\newcommand{\forkIrrelevant}{\textsf{irrelevant}\xspace}
\newcommand{\forkRelevant}{\textsf{relevant}\xspace}
\newcommand{\forkActive}{\textsf{active}\xspace}

\newcommand{\aChain}{\ensuremath{\vec{a}}\xspace}
\newcommand{\hChain}{\ensuremath{\smash[b]{\vec{h}}}\xspace}
\newcommand{\hChainAfter}{\ensuremath{\smash[b]{\vec{h_c}}}\xspace}

\newcommand{\pool}{\textit{pool}\xspace}

\newcommand{\abs}[1]{\ensuremath{\left\lvert{#1}\right\rvert}}
\newcommand{\transactions}[1]{\ensuremath{\#T\left({#1}\right)}}
\newcommand{\emptyChain}{\ensuremath{[~]}}
\newcommand{\increment}{\ensuremath{\texttt{+\!\!\;+}}}

\newcommand{\maxFork}{\ensuremath{L}\xspace}
\newcommand{\maxPool}{\textit{max\_pool}\xspace}

\newcommand{\hap}{\textit{p}\xspace}